\documentclass[11pt]{article}
\usepackage{cite}
\usepackage{amsfonts,amsmath,amssymb,mathrsfs,amsthm}
\usepackage{epsfig,epstopdf,graphicx,graphics}
\usepackage{color}
\usepackage{float}
\usepackage{ifpdf}
\usepackage[colorlinks=true]{hyperref}
\usepackage{braket}

\newtheorem{theorem}{Theorem}[section]

\newtheorem{corollary}{Corollary}[section]

\newtheorem{definition}{Definition}[section]
\newtheorem{example}{Example}[section]

\newtheorem{lemma}{Lemma}[section]

\newtheorem{problem}{Problem}[section]
\newtheorem{proposition}{Proposition}[section]
\newtheorem{remark}{Remark}[section]
\numberwithin{equation}{section}

\newcommand{\bthm}{\begin{theorem}}
\newcommand{\ethm}{\end{theorem}}
\newcommand{\blem}{\begin{lemma}}
\newcommand{\elem}{\end{lemma}}
\newcommand{\bex}{\begin{example}}
\newcommand{\eex}{\end{example}}
\newcommand{\bprop}{\begin{proposition}}
\newcommand{\eprop}{\end{proposition}}
\newcommand{\bplm}{\begin{problem}}
\newcommand{\eplm}{\end{problem}}
\newcommand{\bmrk}{\begin{remark}}
\newcommand{\emrk}{\end{remark}}
\newcommand{\bdfn}{\begin{definition}}
\newcommand{\edfn}{\end{definition}}
\newcommand{\bcor}{\begin{corollary}}
\newcommand{\ecor}{\end{corollary}}

\newcommand{\beq}{\begin{equation}}
\newcommand{\eeq}{\end{equation}}
\newcommand{\beqm}{\begin{equation*}}
\newcommand{\eeqm}{\end{equation*}}
\newcommand{\beqn}{\begin{eqnarray}}
\newcommand{\eeqn}{\end{eqnarray}}
\newcommand{\beqnm}{\begin{eqnarray*}}
\newcommand{\eeqnm}{\end{eqnarray*}}
\newcommand{\bea}{\begin{align}}
\newcommand{\eea}{\end{align}}
\newcommand{\beam}{\begin{align*}}
\newcommand{\eeam}{\end{align*}}

\newcommand{\bei}{\begin{itemize}}
\newcommand{\eei}{\end{itemize}}
\newcommand{\bed}{\begin{description}}
\newcommand{\eed}{\end{description}}
\newcommand{\bee}{\begin{enumerate}}
\newcommand{\eee}{\end{enumerate}}

\newcommand{\bey}{\begin{array}}
\newcommand{\eey}{\end{array}}

\newcommand{\beb}{\begin{thebibliography}{99}}
\newcommand{\eeb}{\end{thebibliography}}

\newcommand{\mbb}{\mathbb}

\newcommand{\la}{\label}

\def\tr{{\rm Tr}}
\def\0{\ket{0}}

\def\l{\langle}
\def\r{\rangle}

\def\ff{\frac}

\def\mm[#1]{{\rm #1}}

\def\tt{\int_{t_0}^t}

\def\inf{\int_{-\infty}^\infty}

\usepackage{scalerel,stackengine}
\stackMath
\newcommand\reallywidehat[1]{%
\savestack{\tmpbox}{\stretchto{%
  \scaleto{%
    \scalerel*[\widthof{\ensuremath{#1}}]{\kern-.6pt\bigwedge\kern-.6pt}%
    {\rule[-\textheight/2]{1ex}{\textheight}}%WIDTH-LIMITED BIG WEDGE
  }{\textheight}% 
}{0.5ex}}%
\stackon[1pt]{#1}{\tmpbox}%
}

\graphicspath{{./pics/}}

\begin{document}

\title{Control of continuous-mode  single-photon states: a review}

\author{Guofeng Zhang\thanks{Department of Applied Mathematics, The Hong Kong Polytechnic University, Hong Kong. Email: guofeng.zhang@polyu.edu.hk.}}

\date{\today}

\maketitle

\begin{abstract}
In this survey, we first introduce quantum fields and open quantum systems, then we present continuous-mode single-photon states and discuss discrete measurements of a single-photon field. After that, we introduce linear quantum systems and show how a linear quantum system responds to a single-photon input. Then we investigate how a coherent feedback network can be used to manipulate the temporal pulse shape of a single-photon state.  Afterwards, we present single-photon filter and master equations.  Finally, we discuss the generation of Schr\"odinger cat states by means of photon addition and subtraction.
\end{abstract}

\tableofcontents

%Things to be added
%\begin{itemize}
%\item photon-catalyzed squeezed vacuum states.

%\item "Vacuum Rabi rate"

%\item single-photon generation in the optical regime  \cite{QPB+15, OOM+16}.  

%\item single-photon generation in the microwave regime  \cite{HSG+07}.  

%\item catching and storing a single photon  \cite{NJY16}

%\item routing a single photon

%\item tomography of a single photon \cite{QPB+15}

%\item a photon can  has a discrete-mode description e.g. in terms of the number of photons, polarization, or a continuous-mode description. 

%\end{itemize}

%\vspace{-3ex}
%%%%%%%%%%%%%%%%%%%%%%%%%%%%%%%%%%%%%
%%%%%%%%%%%%%%%%%%%%%%%%%%%%%%%%%%%%%
%%%%%%%%%%%%%%%%%%%%%%%%%%%%%%%%%%%%%
\section{Introduction}\label{sec:intro}

A light field is said to be in an $\ell$-photon state if it contains {\it exactly} $\ell$ photons. When $\ell=1$, it is in a single-photon state. An optical or microwave photon, as an electromagnetic field, has discrete degrees of freedom like polarization; it also has continuous degrees of freedom, for example a photon can be viewed as a wavepacket with a continuous spectral or spatial envelope.   In this review, we are interested  in continuous-mode single-photon states.  Due to their highly genuine quantum nature, single- and few-photon states hold promising applications in quantum communication, quantum computation, quantum metrology and quantum simulations.  Recently, there has been a rapidly growing interest in the generation, communication, storage, and manipulation  (e.g., pulse shaping) of few-photon states. Thus, a new burgeoning and important problem in the field of quantum engineering is: How to analyze and design quantum dynamical systems driven by few-photon states so as to achieve desirable control performance; for example, producing a single photon with pre-specified pulse shape? In this review, we investigate single-photon states from a control-theoretic perspective. For physical implementation of single photon generation, detection and storing, please refer to  the physics literature \cite{LHA+01,YKS+02,MBB+04,HSG+07,OFV09,BC09,LR09,SFY10,BRV12,PHC+14,LMS,RR15,NJY16,OOM+16,Peng2016,GKM+17,DTK+18,WQD+19,TOS+19} and references therein.

%photon-box

Interaction between a  photon (flying qubit information carrier) and a two-level emitter (stationary qubit information carrier) is fundamental to quantum information processing and quantum physics. Efficient interaction is achieved when the incident photon has a well-defined temporal or frequency modal structure which matches that of the stationary qubit. For example, for atomic excitation by a single photon, it is well-known  \cite{SAL09,WMS+11,PZJ16} that on resonance the optimal excitation by a single photon of a rising exponential pulse shape is achieved when $\gamma=\kappa$, where $\gamma$ is the full width at half maximum (FWHM) of the  photon wave packet and $\kappa$ is the decay rate  of the two-level atom. On the other hand, if the incident photon is of a Gaussian pulse shape, the optimal ratio is $0.8$ which is achieved when $\Omega=1.46\kappa$, where $\Omega$ is the photon frequency bandwidth \cite{SAL09,RSF10,WMS+11,GJN+12,BCB+12}.  When a two-level atom is driven by two co-propagating photons of Gaussian pulse shape, numerical simulations in \cite{SZX16}  show that  the  maximum excitation probability is around $0.88$ attained at  $\Omega=2*1.46\kappa$. Moreover, when a two-level atom is driven by two counter-propagating identical photons, it is shown in \cite{DZA19a,DZA19b}  that the maximum excitation probability is attained at $\gamma=5\kappa$ for rising exponential pulse shapes,  and $\Omega=2*1.46\kappa$ for the Gaussian pulse shapes.

The rest of this article is organized as follows. Open quantum systems are briefly introduced in Section \ref{sec:system}.  Continuous-mode single-photon states are presented in Section \ref{sec:photon_state}. The response of  a quantum linear system to a single-photon input is discussed in Section \ref{sec:linear_resp}.  In Section \ref{sec:shaping} it is shown how to use a  linear coherent feedback network to shape the temporal pulse of a single photon.  Single-photon filters and master equations are  presented in Section \ref{sec:filtering}. Schr\"odinger cat states generation is discussed in Section \ref{sec:cat}. Two possible future research problems are proposed in Section \ref{sec:Con}.

\emph{Notation.} The reduced Planck constant $\hbar$ is set to 1.  $\ket{0}$ stands for the vacuum (namely no photon) state of a free-propagating light field. Given a column vector of operators or complex numbers $X=[x_1,\cdots,x_n]^\top$, the Hilbert space adjoint operator or complex conjugate of $X$ is denoted by $X^\#=[x_1^\ast,\cdots,x_n^\ast]^\top$. Let $X^\dagger=(X^\#)^\top$ and $\breve{X} = [X^\top,  \ X^\dag]^\top$. The commutator between operators $A$ and $B$ is defined to be $[A,B]\triangleq AB-BA$. Given operators $L,H,X,\rho$,  define two superoperators
\begin{eqnarray*}
&\mathrm{Lindbladian}:&\mathcal{L}_GX \triangleq -i[X,H]+\mathcal{D}_LX,\\
&\mathrm{Liouvillian}:&\mathcal{L}_G^\star\rho\triangleq -i[H,\rho]+\mathcal{D}_L^\star\rho,
\end{eqnarray*}
where $\mathcal{D}_LX=L^\dagger XL-\frac{1}{2}(L^\dagger LX+XL^\dagger L)$, and $\mathcal{D}_L^\star\rho=L\rho L^\dagger-\frac{1}{2}(L^\dagger L\rho+\rho L^\dagger L)$.  $\omega_c$ is the system frequency such as the resonance frequency of a cavity mode (which is a quantum harmonic oscillator) or the transition frequency of a two-level system.  $\omega_o$ is the carrier frequency of an external field. $\omega_d = \omega_c - \omega_o$ is the frequency detuning.  $\delta_{jk}$ is the Kronecker delta function and $\delta(t-r)$ is the Dirac delta function.  The Fourier transform of a time-domain function $\xi \in L_{2}(\mathbb{R},\mathbb{C})$ generates
\begin{equation*}\label{xi_nu}
\xi[i\nu ]=\frac{1}{\sqrt{2\pi }}\int_{-\infty}^{\infty } e^{i\nu t}\xi(t) \ dt
\end{equation*}
in the frequency domain,  whose inverse Fourier transform is
\begin{equation}\label{xi_t}
\xi(t)=\frac{1}{\sqrt{2\pi }}\int_{-\infty}^{\infty }\ e^{-i\nu t}\xi[i\nu] \ d\nu.
\end{equation}

%%%%%%%%%%%%%%%%%%%%%%%%%%%%%%%%%%%%%
%%%%%%%%%%%%%%%%%%%%%%%%%%%%%%%%%%%%%
%%%%%%%%%%%%%%%%%%%%%%%%%%%%%%%%%%%%%
\section{Open quantum systems}\label{sec:system}
\begin{figure}[tbph]
\centering
\includegraphics[width=0.6\textwidth]{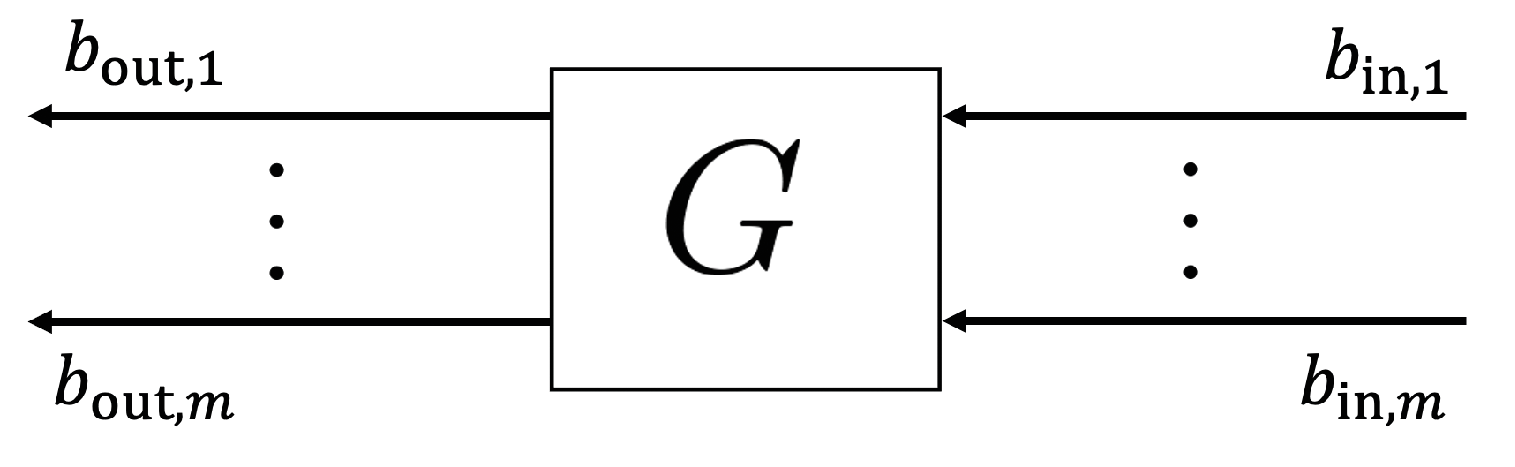}
\caption{A quantum system $G$ with $m$ input fields and $m$ output fields}
\label{fig:sys}
\end{figure}

In this section, we briefly introduce open quantum systems, as shown in Fig. \ref{fig:sys}.  Interested readers may refer to  references \cite{GZ00,BP02,WM09,GJ09,ZJ12,GZ15,CKS17,ZLW+17,NY17,ZGPG18} for mored detailed discussions.

%%%%%%%%%%%%%%%%%%%%%%%%%%%%%%%%%%%%%
%%%%%%%%%%%%%%%%%%%%%%%%%%%%%%%%%%%%%
\subsection{Field}\label{subsec:field}

In quantum physics, a free propagating Boson field is mathematically modeled by a continuum of quantum harmonic oscillators,  represented by their annihilation operators $b_\omega$ and  creation operators $b^\ast_\omega$ (the Hilbert space adjoints of $b_\omega$). Here, $\omega$ stands for frequencies. These field operators satisfy singular commutation relations 
\beq\la{eq:b,feb19}
[b_{\omega}, b_{\omega'}]=[b^\ast_{\omega}, b^\ast_{\omega'}]=0,  \ \ [b_{\omega}, b^\ast_{\omega'}]=\delta(\omega-\omega').
\eeq
Physically,  $b_{\omega}$ annihilates a photon in the field. Hence, if there is no photon in the field, in other words, the field is in the vacuum state $\ket{0}$, then  $b_{\omega}\ket{0}=0$. The Hamiltonian of the field is
\beq \la{eq:may16:Hb}
H_{\mm[B]}= \int_0^\infty d\omega \;  \omega b^\dag_\omega b_\omega.
\eeq
By  \eqref{eq:b,feb19}, in the Heisenberg picture, the free evolution of $b_\omega$ is governed by
\beq\la{eq:may16:b}
b_\omega (t) = e^{i H_{\mm[B]}t}b_\omega e^{-i H_{\mm[B]}t} = e^{-i\omega t} b_\omega.
\eeq 
%Should we mention the initial time for this free evolution ?????
Hence,  
\beq \la{eq:may16:Hb2}
H_{\mm[B]}(t)= \int_0^\infty d\omega \;  \omega b^\dag_\omega(t) b_\omega(t) = H_{\mm[B]},
\eeq
i.e., the Hamiltonian of a free-evolution field is independent of time.

For open quantum systems, we are interested in the interaction between the field and the system of interest,  starting from an initial time. In this paper, we use $t_0$ to denote the initial time. After the initial time, the field does not evolve on its own due to its interaction with the system. As a result,  \eqref{eq:may16:Hb2} does not hold any longer. Instead, the field will carry system's information. It is because of this that the system can be read when the field is measured.

%%%%%%%%%%%%%%%%%%%%%%%%%%%%%%%%%%%%%
\bmrk \la{rem: narrow band approximation}{\rm
A quantum system has a characteristic frequency, e.g., the atomic transition frequency between energy  levels of an atom, the resonant frequency of a quantum cavity resonator, or the Rabi frequency of a continuous-wave (cw) driving laser. When the system interacts with an input field (often called incident field in quantum optics), it is a standard assumption that only the field modes  whose frequencies are near the characteristic frequency of the system contribute to the interaction with the system, while the influence of those  modes whose frequencies are far away from the characteristic frequency of the system is negligible. In other words, it is often the case that the {\it effective} field is within a narrow sideband centered at this characteristic frequency. Under this assumption (narrow band approximation), the range of integral of  \eqref{eq:may16:Hb} can be extended from $-\infty$ to $\infty$.  More discussions can be found in, e.g., \cite{GC85,BLP+90}.
}
\emrk

In the input-output formalism of open quantum systems, the input field in the time domain is  defined as (see e.g. \cite[ (4)]{GC85}, \cite[ (2.14)]{BLP+90}, \cite[Sec. III]{FKS10})
\beq\la{eq:b_in,feb19}
b_\mm[in](t) \triangleq  \ff{1}{\sqrt{2\pi}} \int_{-\infty}^\infty d\omega \; e^{-i\omega(t-t_0)} b_\omega(t_0), \ \ t\geq t_0,
\eeq
where $b_\omega(t_0)$ denotes the quantum harmonic oscillator  $b_\omega$ at the initial time $t_0$.  It can be seen from  \eqref{eq:may16:b} and \eqref{eq:b_in,feb19} that $b_\mm[in](t)$  is a continuum of quantum harmonic oscillators $b_\omega$ for all frequencies. $b^\ast_\mm[in](t)$, the adjoint of $b_\mm[in](t)$, is obtained by conjugating both sides of  \eqref{eq:b_in,feb19}, 
\beqm\la{eq:b_in ast,feb19}
b^\ast_\mm[in](t) =   \ff{1}{\sqrt{2\pi}} \int_{-\infty}^\infty e^{i\omega(t-t_0)} b^\ast_\omega(t_0)d\omega,  \ \ t\geq t_0.
\eeqm
By  \eqref{eq:b,feb19}, we have the singular commutation relations for  $b_\mm[in](t)$ and  $b^\ast_\mm[in](t)$ in the time domain, 
\beq \la{eq:ccr_b t feb19}
[b_\mm[in](t_1), b_\mm[in](t_2)]  = [b^\ast_\mm[in](t_1), b^\ast_\mm[in](t_2)] =0, \ \ [b_\mm[in](t_1), b^\ast_\mm[in](t_2)] = \delta(t_1-t_2), \ \ t_1, t_2\geq t_0.
\eeq
For the  input field $b_\mm[in](t)$ in the time domain defined in   \eqref{eq:b_in,feb19}, we define its Fourier
transform as
\begin{equation}
b_\mm[in][i\omega ]
\triangleq
\frac{1}{\sqrt{2\pi }}\int_{t_0}^{\infty }dt\ e^{i\omega t}b_\mm[in](t),
\ \ \ \omega \in \mathbb{R}.
\label{b_s_1}
\end{equation}
The adjoint $b_\mm[in]^{\ast}[i\omega ]$ of $b_\mm[in][i\omega] $ is obtained by conjugating both sides of  (\ref{b_s_1}); specifically,
\begin{equation*} \label{b_s_2}
b_\mm[in]^{\ast}[i\omega ]
=
\frac{1}{\sqrt{2\pi }}\int_{t_0}^{\infty }dt\ e^{-i\omega t}b_\mm[in]^{\ast}(t),
\ \ \ \omega \in \mathbb{R}.
\end{equation*}
Noticing the identity
\begin{equation*}\label{july2_delta}
\lim_{t_{0}\rightarrow -\infty }\frac{1}{2\pi }\int_{t_{0}}^{\infty }dt\
e^{i\omega t} =\delta (\omega ),
\end{equation*}
it can be readily shown that
\beq \la{eq:may16_delta}
\displaystyle{\lim_{t_{0}\rightarrow -\infty }}[b_\mm[in][i\omega],b_\mm[in]^{\ast}[i\omega']] = \delta(\omega-\omega'), ~~~~ \omega, \omega' \in \mbb{R}.
\eeq

%%%%%%%%%%%%%%%%%%%%%%%%%%%%%%%%%%%%%
\bmrk{\rm
For any fixed $t_0$, applying the inverse Fourier transform to \eqref{eq:b_in,feb19} yields
\beqm
b_\omega(t_0) e^{i\omega t_0} =   \ff{1}{\sqrt{2\pi}} \int_{-\infty}^\infty e^{i\omega t} b_\mm[in](t)   dt, \ \ t\geq t_0.
\eeqm
This, together with \eqref{eq:may16:b}, gives
\beq\la{eq:may16_b_omega}
b_\omega =  \ff{1}{\sqrt{2\pi}} \int_{-\infty}^\infty e^{i\omega t} b_\mm[in](t)   dt, \ \ t\geq t_0.
\eeq
In the limit $t_0\to-\infty$, \eqref{b_s_1} and \eqref{eq:may16_b_omega} yield
\beq\la{eq:may16_b_omega2}
b_\mm[in][i\omega ] = b_\omega,
\eeq
which confirms consistency between \eqref{eq:b,feb19} and \eqref{eq:may16_delta}.
}
\emrk

%%%%%%%%%%%%%%%%%%%%%%%%%%%%%%%%%%%%%
\bmrk{\rm
In the definition of $b_\mm[in](t)$ in  \eqref{eq:b_in,feb19}, if we {\bf implicitly} assume that  $b_\mm[in](t)\equiv0$ for $t<t_0$, then \eqref{b_s_1} becomes
\beq \label{b_s_1_may16}
b_\mm[in][i\omega ]
=
\frac{1}{\sqrt{2\pi }}\int_{-\infty}^{\infty }dt\ e^{i\omega t}b_\mm[in](t),
\ \ \ \omega \in \mathbb{R},
\eeq
which is the same as \eqref{eq:may16_b_omega}. As a result, \eqref{eq:may16_b_omega2} holds without taking the limit $t_0\to-\infty$. Indeed, by \eqref{eq:may16:b} and \eqref{eq:b_in,feb19}, we have 
\beq\la{eq:b_in,may16}
b_\mm[in](t)=  \ff{1}{\sqrt{2\pi}} \int_{-\infty}^\infty e^{-i\omega t} b_\omega d\omega, \ \ t\geq t_0.
\eeq
Thus, under the assumption that  $b_\mm[in](t)\equiv0$ for $t<t_0$, Fourier transforming \eqref{eq:b_in,may16}  yields \eqref{eq:may16_b_omega2}, i.e.,
\beq \label{b_s_1_may16_b} 
b_\omega =
\frac{1}{\sqrt{2\pi }}\int_{-\infty}^{\infty }dt\ e^{i\omega t}b_\mm[in](t) = b_\mm[in][i\omega],
\eeq
where \eqref{b_s_1_may16} has been used. In view of this, it appears meaningful to interpret the variable ``$t$'' in \eqref{eq:b_in,feb19} as the real time under the assumption that $b_\mm[in](t)\equiv0$ for $t<t_0$ where $t_0$ is the time when the system and the field start their interaction. This is fine as we are interested in the dynamics of open quantum systems, instead of free evolution of the system or the field itself. (This interpretation is different from that in \cite[Sec. 2.2]{FTR+18}.)  In the study of stochastic Schr\"{o}dinger  equations it is always assumed that $t_0=0$ or some finite value. On the other hand, in the study of photon scattering off a standing emitter, it is often assumed that $t_0=-\infty$, i.e. the interaction starts in the remote past.  Under the treatment proposed above, the specific value of the initial time $t_0$ is not important. }
%Is it exactly the interaction picture?????
\emrk

We end this subsection as a final remark.

%%%%%%%%%%%%%%%%%%%%%%%%%%%%%%%%%%%%%
\bmrk
In this review, we often omit the subscript ``\;${\mm[in]}$'' for the input fields. Thus $b(t)$ is the input field in the time domain while $b[i\omega]$ is the input field in the frequency domain.  Clearly, $b[i\omega]$ and $b_\omega$ are different objects. However, as shown in \eqref{b_s_1_may16_b}, they can be the same under some condition. 
\emrk

%%%%%%%%%%%%%%%%%%%%%%%%%%%%%%%%%%%%%
%%%%%%%%%%%%%%%%%%%%%%%%%%%%%%%%%%%%%
\subsection{System}\label{subsec:system}

The open Markovian quantum system $G$, as shown in Fig.~\ref{fig:sys},  can be described in the so-called $(S,L,H)$ formalism \cite{GJ09,TNP+11,ZJ12,CKS17}.  
In this formalism, $S,L,H$ are operators on the Hilbert space for the system $G$. Specifically, $S$ is a scattering operator that satisfies $S^\dagger S=SS^\dagger=I$ (the identity operator). For example, $S$ can be a phase shifter or beamsplitter, $L = [L_1,   \ \ldots, \  L_m]^\top$ describes how the system $G$ interacts with its surrounding environment, and the self-adjoint operator $H$ denotes the inherent system Hamiltonian of $G$. The quantum system $G$ is driven by $m$ input fields. Denote the annihilation operator of the $j$-th Boson input field by $b_j(t)$ and the creation operator,  the adjoint operator of  $b_j(t)$,  by $b_j^\ast(t)$, $j=1,\ldots,m$. Then similar to  \eqref{eq:ccr_b t feb19},  the fields satisfy the following singular commutation relations:
\begin{equation} \label{eq:SCR}
\left[b_j(t),b_k^\ast(r)\right]=\delta_{jk}\delta(t-r), \ \ \ j,k=1, \ldots, m, ~{\rm and}~ t, r\geq t_0.
\end{equation}
 Denote a column of vectors $b(t) = [ b_1(t), \cdots,  b_m(t)]^\top$.
The integrated input annihilation, creation, and gauge processes (counting processes) are given by
\begin{eqnarray}\label{eq:gauge}
B(t)=\int_{t_0}^tb(s)ds, ~ B^\#(t)=\int_{t_0}^tb^\#(s)ds, ~ \Lambda(t)=\int_{t_0}^tb^\#(s)b^\top(s)ds,
\end{eqnarray} %$t_0$ or $-\infty$ ?????
respectively. In this paper, the input fields are assumed to be canonical fields which include the vacuum, coherent, single- and multi-photon fields but not the thermal fields.    Then these  quantum stochastic processes satisfy
\begin{eqnarray*}
&&dB_j(t)dB_k^\ast(t)=\delta_{jk}dt, \ \ dB_j(t)d\Lambda_{kl}(t)=\delta_{jk}dB_l(t),
\nonumber
\\
&&d\Lambda_{jk}(t)dB_l^\ast(t)=\delta_{kl}dB_j^\ast(t), \ \ d\Lambda_{jk}(t)d\Lambda_{lm}(t)=\delta_{kl}d\Lambda_{jm}(t).
\label{table}
\end{eqnarray*}

According to quantum mechanics, the whole system in Fig. \ref{fig:sys} evolves in a unitary manner. Specifically, there is a unitary operator $U(t)$ on the tensor product  $\mathrm{System}\otimes\mathrm{Field}$ Hilbert space that governs the temporal evolution of this  quantum system.  It turns out that the unitary operator $U(t)$ is the  solution to the It\^o quantum stochastic differential equation (QSDE)
\begin{equation}\label{QSDE}
dU(t)=\Bigg\{-\left(iH+\frac{1}{2}L^\dagger L\right)dt+LdB^\dagger(t)-L^\dagger SdB(t)+{\rm Tr}[S-I]d\Lambda(t)\Bigg\}U(t)
\end{equation}
with the initial condition $U(t_0)=I$. In particular, if $L=0$ and $S=I$, then (\ref{QSDE}) reduces to 
\begin{equation*}\label{QSDE:Schrodinger}
i\dot{U}=HU,
\end{equation*}
which is the Schr\"{o}dinger equation for an isolated quantum system with Hamiltonian $H$. 

Using the unitary operator $U(t)$ in (\ref{QSDE}), the dynamical evolution of system operators and the environment can be obtained in the Heisenberg picture. Indeed, the time evolution of the system operator $X$, denoted by 
\begin{equation}\label{eq:X}
j_t(X)\equiv X(t)=U^\dagger(t)(X\otimes I_{\mathrm{field}})U(t),
\end{equation}
follows the It\^{o} QSDE
\begin{eqnarray*}
dj_t(X)&=&j_t(\mathcal{L}_GX)dt+dB^\dagger(t)j_t(S^\dagger[X,L])+j_t([L^\dagger,X]S)dB(t)
\nonumber\\
&&+{\rm Tr}[j_t(S^\dagger XS-X)d\Lambda(t)].
\end{eqnarray*}
On the other hand, the dynamical evolution of the output field is given by 
\begin{eqnarray*}
dB_{\mathrm{out}}(t)&=&L(t)dt+S(t)dB(t),
\label{B_out}
\\
d\Lambda_{\mathrm{out}}(t)&=&L^\#(t)L^\top(t)dt+S^\#(t)dB^\#(t)L^\top(t)
\nonumber
\\
&&+L^\#(t)dB^\top(t)S^\top(t)+S^\#(t)d\Lambda(t)S^\top(t),
\label{Lambda_out}
\end{eqnarray*}
where 
\begin{eqnarray*}
B_{\mathrm{out}}(t)&=&U^\dagger(t)(I_{\mathrm{system}}\otimes B(t))U(t),
\\
 \Lambda_{\mathrm{out}}(t)&=&U^\dagger(t)(I_{\mathrm{system}}\otimes \Lambda(t))U(t)
\end{eqnarray*}
 are  the integrated output annihilation operator and gauge processes, respectively.

%%%%%%%%%%%%%%%%%%%%%%%%%%%%%%%%%%%%%
\bex[Optical cavity]\label{ex:cavity}
{\rm Let $G$ be an optical cavity.  Here we consider the simplest case: the cavity has a single intracavity mode (a quantum harmonic oscillator represented by its annihilation operator $a$) which interacts with an external light field represented by its annihilation operator $b(t)$. Because $a$ is a cavity mode, it and its adjoint operator $a^\ast$ satisfy the canonical commutation relation $[a,a^\ast]$=1, in contrast to the singular commutation relation (\ref{eq:SCR}) for free propagating fields.  As introduced in the {\it Notation} part,  let $\omega_d$ be the detuned frequency between the resonance frequency of the internal mode $a$ and the central frequency of the external light field $b(t)$.    Let  $\kappa $ be the half linewidth of the cavity.  In the $(S,L,H)$ formalism, we have
$S=I$, $L=\sqrt{\kappa}a$,  and $H= \omega_d a^\ast a$. Then, the dynamics of this {\it linear} system can be described in the input-output form
\beqn
d a(t) &=& -(i\omega_d+\frac{\kappa}{2}) a(t)dt - \sqrt{\kappa} dB(t) ,
\label{eq: dot a}
\\
dB_{\rm out} (t) &=& \sqrt{\kappa} a(t)dt  + dB(t).
\nonumber
%\label{eq: output}
\eeqn
Applying the Laplace transform to  \eqref{eq: dot a} and omitting the influence of the initial state $a(t_0)$ yield
\beqm
a[i\omega] = -\ff{\sqrt{\kappa}}{i(\omega+\omega_d)+\frac{\kappa}{2}} b[i\omega] =-T[i\omega] b[i\omega],
\eeqm
where
\beq \la{amplitude transmission function}
T[i\omega] \triangleq  \ff{\sqrt{\kappa}}{i(\omega+\omega_d)+\frac{\kappa}{2}}
\eeq
is called the  {\it amplitude transmission function} in optics literature; see e.g., \cite{QPB+15}. $T[i\omega]$ gives the relation between the intra-cavity mode and the input mode and is in the form of a Lorentzian lineshape function; see \eqref{rising_freq}.   Finally, the value $T[0]  = \ff{\sqrt{\kappa}}{i\omega_d+\frac{\kappa}{2}}$ in the neighborhood of  the  resonance of the cavity is commonly used in optics.
}
\eex

%%%%%%%%%%%%%%%%%%%%%%%%%%%%%%%%%%%%%
\bex[Two-level system]\label{ex:atom}
{\rm A two-level system residing in a chiral nanophotonic waveguide can be parametrized by  the triple $
S=1$, $L=\sqrt{\kappa}\sigma_-$, and $H=\frac{\omega_d}{2}\sigma_z $. 
The two-level system has two energy states: the ground state $\ket{g}$ and excited state $\ket{e}$.   $\sigma_- = \ket{g}\bra{e}$ and $\sigma_+ = \ket{e}\bra{g}$ are the lowering and raising operator respectively. We have $\sigma_-\ket{g}=0$ and $\sigma_+\ket{e}=0$.   $\sigma_z = \ket{e}\bra{e}-\ket{g}\bra{g}$ is the Pauli Z operator.  The scalar $\omega_d$ is the detuning frequency between the transition frequency (between $\ket{g}$ and $\ket{e}$) of the two-level system  and the central frequency of the external light field,  and $\kappa$ is the decay rate of the two-level system.   The dynamics of the system are described by}
\begin{eqnarray*}
d\sigma_-(t)&=&-(i\omega_d+\frac{\kappa}{2})\sigma_-(t)dt+\sqrt{\kappa}\sigma_z(t)dB(t) ,
 \label{stwo1}
 \\
dB_{\rm out}(t)&=&\sqrt{\kappa}\sigma_-(t)dt+dB(t).
\label{stwo2}
\end{eqnarray*}
{\rm It should be noted that this system is  {\it bilinear} due to the presence of $\sigma_z(t)dB(t)$, in contrast to the linear resonator in Example \ref{ex:cavity}. However, if the two-level atom is initially in the ground state $\ket{g}$ and the field is in the vacuum state $\ket{0}$, as $\sigma_z(t) \ket{g0} = -\ket{g0}$ (see e.g., \cite[Lemma 3]{PZJ16}), we have
\begin{eqnarray*}
d\sigma_-(t)\ket{g0}&=&-(i\omega_d+\frac{\kappa}{2})\sigma_-(t)dt\ket{g0}-\sqrt{\kappa}dB(t)\ket{g0} ,
 \\
dB_{\rm out}(t)\ket{g0}&=&\sqrt{\kappa}\sigma_-(t)\ket{g0}dt+dB(t)\ket{g0},
\end{eqnarray*}
which is in the form of {\it linear} dynamics.  Finally, rotations
\[ 
\sigma _{-}(t) \to {\rm e}^{{\rm i}\omega _{d}t}\sigma _{-}(t),  \ 
 B(t) \to  {\rm e}^{{\rm i}\omega _{d}t}B(t),  \ B_{\rm out}(t) \to {\rm e}^{{\rm i}\omega _{d}t}B_{\rm out}(t)
\]
put the above system into the following form
\begin{eqnarray*}
d\sigma_-(t)\ket{g0}&=&-\frac{\kappa}{2}\sigma_-(t)dt\ket{g0}-\sqrt{\kappa}dB(t)\ket{g0} ,
 \\
dB_{\rm out}(t)\ket{g0}&=&\sqrt{\kappa}\sigma_-(t)\ket{g0}dt+dB(t)\ket{g0}.
\end{eqnarray*}
}
\eex
%What role is played by $\omega_d$ ?????

%%%%%%%%%%%%%%%%%%%%%%%%%%%%%%%%%%%%%
\bmrk
{\rm In this section, the dynamics of a quantum system are given directly in terms of the system operators $S,L,H$. This is unlike the traditional way  where the starting point is a total Hamiltonian for the joint system-field system, see, e.g, \cite{GC85}. Nevertheless, the $(S,L,H)$ formalism originates from and is a simplified version of the traditional approach \cite{GJ09}.  An illustrative example can be found in \cite[Example 1]{SZX16}.}
\emrk

%%%%%%%%%%%%%%%%%%%%%%%%%%%%%%%%%%%%%
%%%%%%%%%%%%%%%%%%%%%%%%%%%%%%%%%%%%%
%%%%%%%%%%%%%%%%%%%%%%%%%%%%%%%%%%%%%
\section{Continuous-mode single-photon states}\label{sec:photon_state}
In this section, we introduce single-photon states of a free propagating light field.

Denote $\ket{1_t} = b^\ast (t)\ket{0}$. By \eqref{eq:SCR} and $b(t)\ket{0}=0$, we have $\braket{1_t|1_\tau} = \delta(t-\tau)$. A  continuous-mode single-photon state of a light field having the temporal pulse shape $\xi(t) \in L_{2}(\mathbb{R},\mathbb{C})$ can be viewed as a superposition of the continuum of $\ket{1_t}$; in other words,
\beq \la{1 photon jan21}
\ket{1_\xi} \equiv {\bf B}^\ast(\xi)\ket{0} \triangleq \int_{-\infty}^\infty \xi(t) \ket{1_t}d t.
\eeq 
Assume the $L_2$ norm $\|\xi\|\triangleq\sqrt{\int_{-\infty}^\infty |\xi(t)|^2}dt=1$. (It is assumed that $\xi(t)\equiv0$ for $t<t_0$ (the initial time) as we focus on the interaction between a system and a field.)  Then $\braket{1_\xi|1_\xi}=1$. Hence, the probability of finding the photon in the time interval $[t,t+dt)$ is $|\xi(t)|^2 dt$. In the frequency domain, we use $\ket{1_\omega} \triangleq b^\ast [i\omega]\ket{0}$.  In the frequency domain \eqref{1 photon jan21} becomes
\beq \la{1 photon jan28}
\ket{1_\xi} = \int_{-\infty}^\infty \xi[i\omega] \ket{1_\omega}d \omega.
\eeq

%%%%%%%%%%%%%%%%%%%%%%%%%%%%%%%%%%%%%
\bmrk{\rm
If we do not restrict $\xi$ to be in $L_{2}(\mathbb{R},\mathbb{C})$, for example, let $\xi[i\omega] = \delta(\omega-\omega_0)$ for some real $\omega_0$; in other words, we have a monochromatic light field with frequency $\omega_0$. Then from \eqref{1 photon jan28} we  get
\beqm \la{1 photon jan29}
\ket{1_\xi} =  b^\ast[i\omega_0]\ket{0} = \ket{1_{\omega_0}}.
\eeqm
 Moreover, by  \eqref{xi_t} we get  the temporal wave packet $\xi(t) = \ff{1}{\sqrt{2\pi}}e^{i\omega_0 t}$ whose modulus is $|\xi(t)|\equiv  \ff{1}{\sqrt{2\pi}}$ for all $t\in \mathbb{R}$. 
 }
\emrk

%\bplm
%For the monochromatic light field above-mentioned, $\braket{1_\xi|1_\xi}\neq1$.  What is the a light field with temporal wavepacket $\xi(t) = \ff{1}{\sqrt{2\pi}}e^{i\omega_0 t}$? At least, its state is not a continuous-mode single-photon state.  ?????
%\eplm 
%
%
%If $\xi(t) = \delta(t-T)$ for some real $T$, then from \eqref{1 photon jan21} we get
%\beq
%\ket{1_\xi}  = \ket{1_T}.
%\eeq
%Moreover, $\xi[i\omega] = \ff{1}{\sqrt{2\pi}}e^{-i\omega T}$.  Clearly, $|\xi[i\omega]|\equiv \ff{1}{\sqrt{2\pi}}$ for all $\omega\in \mathbb{R}$. Moreover, by \eqref{1 photon jan28} we have in the frequency domain 
%\beq \la{2 photon jan28}
%\ket{1_\xi} =  \ff{1}{\sqrt{2\pi}}\int_{-\infty}^\infty e^{-i\omega T} b^\ast[i\omega] d\omega \ket{0}.
%\eeq

If a field is restricted to be in the interval  $[t_0,t]$,  \eqref{1 photon jan21} becomes
\beqm \la{1 photon 2 jan21}
\ket{1_\xi} = \tt \xi(r) \ket{1_r}d r.
\eeqm 
Similarly, an $\ell$-photon state of a field over the interval $[t_0,t]$  can be defined as
\beq \la{l photon jan21}
\ket{\ell_\psi} =\tt \cdots \tt \psi(\tau_1,\ldots,\tau_\ell)\ket{1_{\tau_1}}\cdots \ket{1_{\tau_\ell}}d\tau_1 \cdots \ d\tau_\ell.
\eeq
In other words, by means of a continuum of basis vectors $\{\ket{0}, \ket{1_{\tau_1}},\cdots, \ket{1_{\tau_\ell}}: \tau_1,\ldots,\tau_\ell \in [t_0, t] \}$, any  $\ell$-photon state can be expressed in the way given in  \eqref{l photon jan21}. As photons are indistinguishable, the function $\psi$ in  \eqref{l photon jan21} is permutation-invariant w.r.t. $\tau_1,\ldots,\tau_\ell$.

  Under the single-photon state $\ket{1_\xi} $, the field operator $b(t)$, which is a quantum stochastic process, has  zero mean, and whose covariance function is
\begin{eqnarray*}\label{eq:R-photon}
R(t,r)\triangleq
\langle 1_\xi | \breve b(t) \breve b^\dag(r)| 1_\xi\rangle
=\delta(t-r) \left[\begin{array}{ll}
                     1 & 0 \\
                     0 & 0
                   \end{array}
 \right] +
\left[ \begin{array}{ll}
 \xi(r)^\ast \xi(t)  &  0
\\
0 & \xi(r)\xi(t)^\ast
\end{array} \right], \ \ t,r\geq t_0.
\end{eqnarray*}

By  (\ref{eq:gauge}), the gauge process is $\Lambda(t) = \int_{t_0}^t n(r) dr$, where $n(t) \triangleq b^\ast(t) b(t)$ is the number operator for the field. In the case of the single-photon state $\ket{1_\xi}$ defined in  \eqref{1 photon jan21}, the intensity is the mean $\bar n(t) \triangleq \langle 1_\xi \vert n(t) \vert 1_\xi \rangle = \vert \xi(t) \vert^2$. Clearly, $\int_{t_0}^\infty \bar n(t)dt=1$, i.e., there is one photon in the field.

Next, we look at three commonly used single-photon states. Firstly, when $\xi(t)$  is an exponentially decaying pulse shape
\begin{equation}\label{31}
\xi(t)=\left\{\begin{array}{ll}
                \sqrt{\beta}e^{-\left(\ff{\beta}{2}-i\omega_o\right) t}, & t\geq0, \\
                0, & t<0,
              \end{array}\right.
\end{equation}
 the state $\ket{1_\xi}$ can describe a single photon  emitted from an optical cavity with resonant frequency $\omega_o$ and  damping rate $\beta$ or a two-level atom with atomic transition frequency $\omega_o$  and decay rate $\beta$  \cite{WM08,RL00}.  The frequency counterpart of  \eqref{31} is
\beqm \la{decaying_freq}
\xi[i\omega] = \sqrt{\frac{\beta}{2\pi}}\frac{1}{\frac{\beta}{2}+i(\omega-\omega_o)}.
\eeqm
Secondly, if  $\xi(t)$  is a rising exponential pulse shape 
\begin{equation}\label{50}
\xi(t)=\left\{\begin{array}{ll}
                \sqrt{\beta}e^{\left(\ff{\beta}{2}+i\omega_o\right) t}, & t\leq0, \\
                0, & t>0,
              \end{array}\right.
\end{equation}
in the frequency representation it is
\beq \la{rising_freq}
\xi[i\omega] = \sqrt{\frac{\beta}{2\pi}}\frac{1}{\frac{\beta}{2}-i(\omega-\omega_o)}.
\eeq
where $\omega_o$ is the carrier frequency of the light field, then on resonance the single-photon state  $\ket{1_\xi}$ is able to fully excite a two-level system if $\beta = \kappa$, where $\kappa$ is the decay rate as introduced  in Example \ref{ex:atom}, see, e.g., \cite{SAL09,WMS+11,YJ14,PZJ16}.  The single photon with pulse shape (\ref{31}) or (\ref{50}) has Lorentzian lineshape function with the full width at half maximum (FWHM) $\beta$ \cite{RL00,BR04}, which in the {\it frequency domain} satisfies
\beqm\la{eq:apr11_freq}
|\xi[i\omega]|^2 = \ff{1}{2\pi}\ff{\beta}{(\omega-\omega_o)^2 + \left(\ff{\beta}{2}\right)^2},
\eeqm
which is \cite[ (3.28)]{BLP+90} with $F=1$ (the single-photon case). 

Finally, the Gaussian pulse shape can be given by
\begin{equation} \label{51}
\xi(t)=\left(\frac{\Omega^2}{2\pi}\right)^{\frac{1}{4}}\exp\left(-\frac{\Omega^2}{4}(t-\tau)^2\right),
\end{equation}
where $\tau$ is the photon peak arrival time. Applying Fourier transform to $\xi(t)$ in   (\ref{51}) we get 
\beqm\la{eq:apr11_freq_Gauss}
|\xi[i\omega]|^2 =\frac{ 1}{\sqrt{2\pi}\; (\Omega/2) }\exp \left(-\frac{\omega ^2}{2(\Omega/2) ^2}\right).
\eeqm
Hence, $\Omega$ is the {\rm frequency} bandwidth of the single-photon wavepacket. Actually,
\beqnm
\xi[i\omega] 
&=& 
\ff{1}{(\ff{\pi}{2}\Omega^2)^{1/4}}\exp\left(\omega(i\tau - \ff{\omega}{\Omega^2}))\right)
\nonumber
\\
&=& 
\ff{1}{(\ff{\pi}{2}\Omega^2)^{1/4}} \exp\left(-\bigg(\ff{\omega-i\pi\Omega/2}{\Omega}\bigg)^2 - \bigg(\ff{\tau\Omega}{2}\bigg)^2\right).
\label{eq:Gaussian_omega}
\eeqnm
Let $\tau=0$ and $\Omega = \sqrt{2}R$. Then \eqref{eq:Gaussian_omega} becomes
\beq\la{eq:gaussian_omega}
\xi[i\omega]  = \ff{1}{\pi^{1/4}\sqrt{R}} \exp\left(-\ff{(\omega/R)^2}{2}\right).
\eeq
Similarly, \eqref{51} becomes
\beq\la{eq:gaussian_t}
\xi(t)=\ff{\sqrt{R}}{\pi^{1/4}}\exp\left(-\ff{(Rt)^2}{2}\right).
\eeq
\eqref{eq:gaussian_omega}-\eqref{eq:gaussian_t} are the form of the wavefunction of a vacuum state \cite[Chapter 4]{GZ00}. In particular, if $R\neq 1$, \eqref{eq:gaussian_omega}-\eqref{eq:gaussian_t} are in the form of the wavefunction of a squeezed vacuum state \cite[Eqs. (5.3)-(5.4)]{AL15}. This is not surprising because vacuum states and squeezed vacuum states are coherent states which have Gaussian wave-packets. 

%Wigner spectrum ?????

%%%%%%%%%%%%%%%%%%%%%%%%%%%%%%%%%%%%%%
%\bmrk
%{\rm It is worth noticing that the frequency bandwidth $\Omega$ is of the from $(\omega_o-\Omega,\omega_o+\Omega)$.} %?????
%\emrk

In contrast to the full excitation of a two-level atom by a single photon of rising exponential pulse shape \eqref{50}, the maximal excitation probability of a two-level atom by a single photon of Gaussian pulse shape \eqref{51} is around 0.8 which is achieved at $\Omega = 1.46\kappa$, see, e.g., \cite{SAL09,WMS+11,YJ14,SZX16,PZJ16}. 

For a mathematical theory of how to generate a single-photon of a prescribed temporal pulse, interested readers may refer to \cite{GZ15b}. For physical generation of a single photon of rising exponential pulse shape, interested readers may refer to \cite{QPB+15,OOM+16}.

%%%%%%%%%%%%%%%%%%%%%%%%%%%%%%%%%%%%%
\bmrk
{\rm It should be noted that a continuous-mode single-photon state $\ket{1_\xi}$ discussed above is different from a continuous-mode single-photon {\it coherent} state $\ket{\alpha_\xi}$ which can be defined as
\beq \la{eq:may21_coherent_1}
\ket{\alpha_\xi} \triangleq \exp(\alpha {\bf B}^\ast(\xi) -\alpha^\ast {\bf B}(\xi) )\ket{0},
\eeq
where $\alpha = e^{i\theta}\in \mathbb{C}$.  For $\ket{\alpha_\xi}$, although the mean photon number is   
\[
\l \alpha_\xi |  {\bf B}^\ast(\xi){\bf B}(\xi) |\alpha_\xi \r = |\alpha|^2=1,
\] which is the same as the single-photon state $\ket{1_\xi}$,  the mean amplitude is $\l \alpha_\xi |  {\bf B}(\xi)| \alpha_\xi\r = \alpha$. In contrast, the mean amplitude  of the  single-photon state $\ket{1_\xi}$  is $\l 1_\xi |  {\bf B}(\xi)| 1_\xi\r =0$. More discussions can be found in \cite[Section 2.1]{DZA16}, \cite[Section 7.1.1]{CKS17}. Finally, continuum coherent states are defined in \cite[ (3.1) and (3.6)]{BLP+90} and \cite[(24)]{WMS+11}, which in our notation are of the form
\beq\la{eq:may21_BLP+3.1}
\ket{\{\alpha[i\omega]\}} = \exp\left(\inf d\omega\;  (\alpha[i\omega] b[i\omega]^\dag -  \alpha[i\omega]^\ast b[i\omega])\right)\ket{0}.
\eeq
Clearly, if $\alpha[i\omega] = \alpha \xi[i\omega]$,  \eqref{eq:may21_BLP+3.1} becomes  \eqref{eq:may21_coherent_1}. A time-domain discussion can be found in \cite{YJ14}.} 
\emrk

If an electromagnetic field is confined for example in a cavity, it will have discrete modes instead of a continuum of modes. Next, we introduce {\it single-mode} coherent states. If a laser is incident on the cavity, the cavity can be in a coherent state.  Moreover, in quantum optics, laser is often assumed to produce a single-mode coherent signal. The reason is simple, if the pulse $\xi[i\omega] \equiv\gamma\delta[\omega_o]$ in \eqref{eq:may21_coherent_1}, where $\gamma\in\mathbb{R}$ and $\omega_o$ is the central frequency,  then we have a {\it single-mode} coherent state
\beq \la{eq:coherent state 0}
\ket{\beta}  = \exp(\beta b^\ast[i\omega_o] -\beta^\ast b[i\omega_o])\ket{0},
\eeq
 where $\beta = \alpha  \gamma \in \mathbb{C}$. The single-mode coherent state $\ket{\beta}$  can be rewritten as
\beq\label{eq:coherent state}
\ket{\beta} = e^{-\ff{1}{2}|\beta|^2} \sum_{n=0}^\infty \ff{\beta^n}{\sqrt{n!}}\ket{n}.
\eeq
Clearly, the vacuum state $\ket{0}$ is a coherent state ($\beta=0$ in \eqref{eq:coherent state 0}). Another type of single-mode coherent states, single-mode squeezed vacuum states, will be introduced in Section \ref{sec:cat}.  Finally, it is worthwhile to point out that there is a slight abuse of notation in \eqref{eq:coherent state 0}. In this survey, the annihilation operator of a free propagating field is denoted by $b(t)$, while that for a cavity is denoted by $a(t)$.  However, as we want to show that the single-mode coherent state $\ket{\beta}$ in \eqref{eq:coherent state 0}  is an ideal approximation of the continuous-mode coherent state  $\ket{\alpha_\xi}$ in \eqref{eq:may21_coherent_1}, we feel it is good to use $b(t)$ in both of these two equations.
 
More discussions on coherent states can be found in \cite[Chapter 4]{GZ00}.

%Plot Wigner functions for a single-photon Fock state and a single-photon coherent state ?????

%%%%%%%%%%%%%%%%%%%%%%%%%%%%%%%%%%%%%
\bmrk
{\rm
The operator ${\bf B}^\ast(\xi)$ defined via  \eqref{1 photon jan21} is called a discrete photon creation operator in the temporal modes theory of quantum optics; see e.g.,  \cite[ (2.14)]{TG66} and  \cite[ (7)]{RW20}. Roughly speaking,  under the {\it narrow-band approximation} as described  in Remark \ref{rem: narrow band approximation},  using $b[i\omega]$ we can define a time-spatial operator, \cite[ (7.3)]{WM08},
\beqm \la{eq:may21_a_x_t}
b(x,t) = e^{-i\omega_o(t-x/c)} \ff{1}{\sqrt{2\pi}} \inf d \omega' \; b[i\omega'] e^{-i\omega'(t-x/c)}, 
\eeqm
where $\omega_o$ is the carrier frequency of the free propagating field.  Denote
\beqn
\mathcal{E}^+(x,t) &=& i b(x,t), \label{eq:E+} \\
\mathcal{E}^-(x,t)  &=& -ib(x,t)^\ast = (\mathcal{E}^+(x,t))^\ast. \nonumber
\eeqn
 Then in a simplified form, an electromagnetic field propagating along the positive $x$ direction can be described as, \cite{milburn08},  
\beqm
\mathcal{E}(x,t) = \mathcal{E}^+(x,t) + \mathcal{E}^-(x,t) =
i(b(x,t) - b(x,t)^\ast).
\eeqm
Let $\{\xi_j\}$ be an orthonormal basis of the space of the square-integrable pulse shapes $\xi$. An example of $\{\xi_j[i\omega]\}$  is the set of wighted Hermite polynomials \cite[Chapter 8]{Hassani13}.   Define 
\beqm
{\bf B}_j \triangleq {\bf B}(\xi_j) = \inf dt\; b(t)\xi_j^\ast(t) = \inf d\omega\; b[i\omega] \xi_j^\ast[i\omega].
\eeqm
Clearly, ${\bf B}_j^\ast\ket{0}$ generates a photon of pulse shape $\xi_j$. Moreover, as  $\{\xi_j\}$ is an orthonormal basis, we have
\beq \label{eq:b and B}
\sum_j \xi_j[i\omega] {\bf B}_j  = \inf   d\omega'\; b[i\omega'] \sum_j \xi_j[i\omega] \xi_j^\ast[i\omega']  \inf   d\omega'\; b[i\omega'] \delta (\omega-\omega')= b[i\omega].
\eeq
Define a time-spatial function $\nu_j(x,t)$ which is associated with $\xi_j$ by
\beq \label{eq:nu}
\nu_j(x,t) \triangleq  i e^{-i\omega_o(t-x/c)} \ff{1}{\sqrt{2\pi}} \inf d \omega' \; \xi_j[i\omega'] e^{-i\omega'(t-x/c)}.
\eeq
By  \eqref{eq:b and B}-\eqref{eq:nu}, the positive-frequency part $\mathcal{E}^+(x,t)$ of an electromagnetic field defined in  \eqref{eq:E+}  can be re-written as
\beqm
\mathcal{E}^+(x,t) = \sum_j  \nu_j(x,t) {\bf B}_j.
\eeqm
More discussions can be found in a recent review on temporal modes in quantum optics \cite{RW20}.
}
\emrk

%%%%%%%%%%%%%%%%%%%%%%%%%%%%%%%%%%%%%
%%%%%%%%%%%%%%%%%%%%%%%%%%%%%%%%%%%%%
\subsection{Discrete measurement of a continuous-mode single-photon  state} \label{subsec:digital}

In this subsection,  we present a procedure of digitizing  a continuous-mode single-photon  state $\ket{1_\xi}$ as is proposed in \cite{QPB+15}. 

Assume that the time interval $(-\infty,\infty)$ is partitioned into  time bins of equal length $\Delta t$ and let  $t_j = j \Delta t$ for $j=0,\pm1,\ldots$.  When $\Delta t$ is sufficiently small, the value of $\xi(t)$ in the time bin $[j\Delta t, (j+1)\Delta t)$ is well approximated by $\xi_{t_j}$. Then a continuous-mode single-photon state $\ket{1_\xi}$ can be approximately written as 
\beq\la{eq:discrete}
\ket{1_\xi} = \int_{-\infty}^\infty \xi^\ast(t)b^\ast(t)dt\0 = \sum_j \xi^\ast(t_j)\frac{1}{\Delta t}\int_{t_{j}}^{t_{j+1}} b^\ast(t)dt\0. 
\eeq
Denote
\beq\la{eq:1_j}
\ket{1_j} =   c^\ast_j \0 \equiv \frac{1}{\Delta t} \int_{t_{j}}^{t_{j+1}} b^\ast(t)dt\0 .
\eeq
Notice that
\beq\la{CCR}
[c_j, c_k^\ast] = \ff{\delta_{jk}}{\Delta t}. 
\eeq

%%%%%%%%%%%%%%%%%%%%%%%%%%%%%%%%%%%
\bmrk{\rm
According to  \eqref{CCR},  $[c_j, c_j^\ast] $ becomes a Dirac delta function $\delta (t_j)$ in the limit $\Delta t\to 0$. This justifies the coefficient    $\frac{1}{\Delta t}$ in the definition of $c_j^\ast$ in  \eqref{eq:1_j}.
}
\emrk

By means of  \eqref{eq:1_j}, the approximated single-photon state in \eqref{eq:discrete} can be re-written as
\beqm\la{eq:discrete2}
\ket{1_\xi}  =  \sum_j \xi^\ast(t_j)\ket{1_j}.
\eeqm
The corresponding density matrix is
\beq\la{eq:density}
\rho = \ket{1_\xi} \bra{1_\xi}=\sum_{mn} \rho_{mn}\ket{1_m}\bra{1_n},
\eeq
where $\rho_{mn} = \xi^\ast(t_m)\xi(t_n)$.  Define a quadrature $X_j$ for the $j$th time bin to be
\beq \la{eq X_j}
X_j \triangleq \frac{c_j e^{-i\theta_j}+c_j^\ast e^{i\theta_j}}{\sqrt{2}}
\eeq
where
\beqm
 \theta_j = \omega_d\cdot t_j + \theta_0
\eeqm
with $\omega_d$ being the frequency  detuning between the central frequency of  the signal (the single-photon state $\ket{1_\xi}$) and that of the local oscillator for homodyne measurement, and $\theta_0$ being the local oscillator's relative phase at $t=0$. As $X_j=X_j^\ast$, $X_j$ is an observable which can be measured in principle. Moreover, it can be shown that 
\beqm
[X_j,\ X_k] =-\ff{i\delta_{jk}}{\Delta t} \sin ((j-k)\omega_d \Delta t) =0, ~~~ j\neq k.
\eeqm
The time series $\{X_j\}$ can be measured in experiments. 

%%%%%%%%%%%%%%%%%%%%%%%%%%%%%%%%%%%
\bmrk{\rm
Notice that  the quadrature $X_j$ defined in  \eqref{eq X_j} can be re-written as
\beqm
X_j  = \ff{c_j +c_j^\ast}{\sqrt{2}}\cos\theta_j +  \ff{c_j -c_j^\ast}{i\sqrt{2}}\sin\theta_j \equiv Q_j \cos\theta_j + P_j \sin\theta_j,
\eeqm
where the quadratures $Q_j = \ff{c_j +c_j^\ast}{\sqrt{2}}$ and $P_j =  \ff{c_j -c_j^\ast}{i\sqrt{2}}$ satisfy
\beqm
[Q_j, P_j] = \ff{i}{\Delta t}.
\eeqm
}
\emrk

It can be calculated that 
\beq \la{eq: X_k 1_m}
X_k \ket{1_m} = \frac{e^{-i\theta_k}}{\sqrt{2}} \delta_{km} \ket{0} +\ff{1}{\Delta t} \frac{e^{i\theta_k}}{\sqrt{2}}\int_{t_j}^{t_{j+1}} dt \int_{t_m}^{t_{m+1}} dr b^\ast (t) b^\ast (r)\0.
\eeq
Therefore, we have
\beqn\la{eq:may22_1}
\l 1_n |X_j X_k|1_m \r = \frac{1}{2} (e^{-i(\theta_k-\theta_j)}\delta_{km}\delta_{jn}+\delta_{jk}\delta_{mn}+ e^{-i(\theta_j-\theta_k)}\delta_{jm}\delta_{kn}).
\eeqn
Let $I(t_j)$ be the homodyne current for the $j$th time bin, which is proportional to $X_j$.  By  \eqref{eq:density} and \eqref{eq:may22_1}, we have
\beqnm 
&&\l I(t_j)I(t_k) \r \propto \l X_j X_k \r
\\
 &=& \tr[\rho X_j X_k]  = \sum_{mn}\rho_{mn}\l 1_n |X_j X_k|1_m \r 
\nonumber
\\  
&=& \frac{1}{2} (e^{-i(\theta_k-\theta_j)}\rho_{kj} + \delta_{jk}+e^{-i(\theta_j-\theta_k)} \rho_{jk})
\nonumber
\\
&=&
\frac{1}{2}\delta_{jk}+{\rm Re}[\rho_{jk}]\cos(\theta_j-\theta_k)+{\rm Im}[\rho_{jk}]\sin(\theta_j-\theta_k)
\nonumber
\\
&=&
\frac{1}{2}\delta_{jk}+{\rm Re}[\rho_{jk}]\cos(\omega_d (t_j-t_k))+{\rm Im}[\rho_{jk}]\sin(\omega_d (t_j-t_k)),
\la{eq: mean X_j X_k}
\eeqnm
which is \cite[ (5)]{QPB+15}.

%%%%%%%%%%%%%%%%%%%%%%%%%%%%%%%%%%%
\bmrk{\rm
In an experiment, it is the homodyne photocurrent $I(t)$ that is recorded, from which $ \l  X_j X_k \r $ is obtained by averaging over many trajectories. After getting $ \l  X_j X_k \r $, ${\rm Re}[\rho_{jk}]$ and ${\rm Im}[\rho_{jk}]$  can be retrieved, from which we can get a discrete approximation of  $\xi(t)$.
}
\emrk

%%%%%%%%%%%%%%%%%%%%%%%%%%%%%%%%%%%%%
%%%%%%%%%%%%%%%%%%%%%%%%%%%%%%%%%%%%%
%%%%%%%%%%%%%%%%%%%%%%%%%%%%%%%%%%%%%
\section{Linear systems' response to single-photon states} \label{sec:linear_resp}

Let the system $G$ in Fig.~\ref{fig:sys} be linear and driven by $m$ photons, one in each input field. Also, assume that $G$ is initialized in a coherent state. Single-mode coherent states are defined in \eqref{eq:coherent state}, its multi-mode counterpart can be found in  \cite[Section II-E]{ZJ13}.  In this section, we present the state of the output fields.

Given two constant matrices $U$, $V\in \mathbb{C}^{r\times k}$, a doubled-up matrix $\Delta\left(U,V\right) $ is defined as
\begin{equation*}
\Delta\left(U,V\right)\triangleq\left[
\begin{array}{ll}
U & V \\
V^{\#} & U^{\#}
\end{array}
\right] .
\end{equation*}
Let $I_{k}$ be an identity matrix and $0_k$ a zero square matrix, both of dimension $k$. Define $J_{k}=\mathrm{diag}%
(I_{k},-I_{k})$. Then for a matrix $X\in\mathbb{C}^{2j\times 2k}$, define $X^{\flat}=J_{k}X^{\dag}J_{j}$. In the linear case, the system $G$ can be used to model a collection of $n$ quantum harmonic oscillators that are driven by $m$ input fields.  Denote $a(t) = [a_1(t) \ \ \cdots \ \  a_n(t)]^\top$, where $a_j(t)$ is the annihilation operator for the $j$th harmonic oscillator, $j=1,\ldots, n$.  In the $(S,L,H)$ formalism,  the inherent
system Hamiltonian is given by
$H=(1/2)\breve{a}^{\dag }\Omega \breve{a}$, where
$a
=
[a^{\top } \; (a^{\#})^{\top}]^{\top }$, and
$\Omega =\Delta (\Omega _{-},\Omega _{+})\in \mathbb{C}^{2n\times 2n}$
is a Hermitian matrix with $\Omega _{-},\Omega _{+}\in
\mathbb{C}^{n\times n}$. The coupling between the system and the fields is
described by the operator
$L=[C_{-} \ C_{+}]\breve{a}$,
with $C_{-},C_{+}\in \mathbb{C}^{m\times n}$.  Finally,  the scattering operator $S$ is an $m\times m$ constant  matrix such that $S^\dag S = S S^\dag = I_m$. The dynamics of the open quantum linear system in Fig.~\ref{fig:sys} are
described by the following linear It\^{o} QSDEs (\cite[ (26)]{GJN10a}, \cite[ (14)-(15)]{ZJ13}, \cite[ (5)-(6)]{ZGPG18}, \cite{ZJ11,ZPL20}, \cite[Chapter 2]{NY17}), 
\begin{equation}\label{eq:sys_a}
\begin{split}
d\breve{a}(t)
=&\; Aa(t)dt+Bd\breve{B}(t),
 \\
d\breve{B}_{\mathrm{out}}(t)
=&\; Ca(t)dt+Dd\breve{B}(t),
\ \ t\geq t_0,
\end{split}
\end{equation}
where 
\[
 D =\Delta\left(S,0\right),   C=\Delta\left( C_{-},C_{+}\right), B=-C^{\flat}\Delta\left(S,0\right), 
 A=-\frac{1}{2}C^{\flat}C-iJ_{n}\Delta\left(\Omega_{-},\Omega_{+}\right).
\]
 These constant  system matrices are parametrized by the physical parameters $\Omega_{-},\Omega_{+},C_{-},C_{+}$ and satisfy
\begin{subequations}
\beqn
A+A^{\flat }+BB^{\flat } &=& 0, 
 \label{eq:PR_a}
\\
B &=&-C^{\flat }\Delta\left(S,0\right).
 \label{eq:PR_b}
\eeqn
\end{subequations}
%\beq
%[\breve{a}, \breve{a}^\dag] \triangleq \breve{a}\breve{a}^\dag - (\breve{a}^\#\breve{a}^\top)^\top,
%\eeq
  (\ref{eq:PR_a}) is equivalent to
\beqm
[\breve{a}(t), \breve{a}^\dag(t)] \equiv [\breve{a}(t_0), \breve{a}^\dag(t_0)] = J_n,  \ \ \forall t\geq t_0.
\eeqm 
That is, the system variables preserve canonical commutation relations. On the other hand, 
  (\ref{eq:PR_b}) is equivalent to
\beqm
[\breve{a}(t), \breve{b}_{\rm out}^\dag (r)]=0, \ \ t\geq r\geq t_0.
\eeqm
That is, the system variables and the output satisfy the non-demolition condition.  In the quantum control literature, equations (\ref{eq:PR_a})-(\ref{eq:PR_b}) are called {\it physical realization} conditions. Roughly speaking, if these conditions are met, the mathematical model \eqref{eq:sys_a}   could in principle be physically realized (\cite{JNP08}, \cite{NJD09}).

Let $X$ be an operator on the joint system-field space. Denote by $\left\langle X(t)\right\rangle $ the
expected value of $X(t)$ with respect to the initial joint system-field state.
Then  \eqref{eq:sys_a} has a corresponding {\it classical} linear system of the form
\begin{equation}\label{eq:sys_a_classical}
\begin{split}
d \l \breve{a}(t) \r 
=&\; A \l a(t) \r dt+Bd \l \breve{B}(t)\r, 
 \\
d\l \breve{B}_{\mathrm{out}}(t)\r
=&\; C \l a(t) \r dt+Dd\l \breve{B}(t) \r,
\ \ t\geq t_0,
\end{split}
\end{equation}
By means of the classical  linear systems theory, we can define Hurwitz stability, controllability and observability of a linear quantum system.

%%%%%%%%%%%%%%%%%%%%%%%%%%%%%%%%%%%%%
\begin{definition} \cite[Definition 1]{GZ15} \label{def:stab_ctrb_obsv} 
The quantum linear system \eqref{eq:sys_a} is said to be \textit{Hurwitz stable} (resp. \textit{controllable}, \textit{observable}) if the corresponding classical linear system \eqref{eq:sys_a_classical} is \textit{Hurwitz stable} (resp. \textit{controllable}, \textit{observable}).
\end{definition}

Moreover, as in classical linear systems theory, the \emph{impulse response function} for the system $G$ is defined as
\begin{equation}\label{eq:gg}
g_{G}(t)\triangleq\left\{
\begin{array}{ll}
\delta(t)D-Ce^{At}C^{\flat}D, & t\geq 0,   \\
0, & t<0.
\end{array}
\right. 
\end{equation}
It is easy to show that $g_{G}(t)$ defined in (\ref{eq:gg}) has the following nice structure
\begin{equation*}
g_{G}(t)=\Delta\left( g_{G^{-}}(t),g_{G^{+}}(t)\right),  \label{eq:impulse}
\end{equation*}
where
\begin{eqnarray*}
g_{G^{-}}(t)&\triangleq&\left\{
\begin{array}{ll}
\delta(t)S-[%
\begin{array}{ll}
C_{-} & C_{+}%
\end{array}
]e^{At}\left[
\begin{array}{c}
C_{-}^{\dag} \\
-C_{+}^{\dag}%
\end{array}
\right]S, & t\geq 0   \\
0, & t<0 %
\end{array}
\right., \nonumber \\
\ g_{G^{+}}(t)&\triangleq&\left\{
\begin{array}{ll}
-[%
\begin{array}{ll}
C_{-} & C_{+}%
\end{array}
]e^{At}\left[
\begin{array}{c}
-C_{+}^{T} \\
C_{-}^{T}%
\end{array}
\right] , & t\geq 0  \\
0 , & t<0  %
\end{array}
\right. .  \label{eq:io}
\end{eqnarray*}
Given a function $f(t)$ in the time domain, its two-sided Laplace transform  \cite[Chapter 10]{WRL61} is defined as
\begin{equation*}
F[s] \equiv \mathscr{L}_b \{f(t)\}(s)  \triangleq \int_{-\infty}^\infty e^{-st} f(t) dt.
\end{equation*}
Applying the two-sided Laplace transform to the impulse response function (\ref{eq:gg}) yields the transfer function
\begin{equation*}\label{eq:G_omega}
\Xi_G[s] = \Delta(  \Xi_{G^-}[s], \Xi_{G^+}[s] ),
\end{equation*}
where $\Xi_{G^-}[s]= \mathscr{L}_b \{g_{G^-}(t)\}(s)$ and $\Xi_{G^+}[s]= \mathscr{L}_b \{g_{G^+}(t)\}(s)$.

If $C^+=0$ and $\Omega^+=0$, the resulting quantum linear system is said to be {\it passive} \cite{ZJ11,GZ15,ZGPG18}. Specifically, the It\^o QSDEs for a passive linear
quantum system are (e.g., see \cite[Sec. 3.1]{GZ15}), 
\begin{equation}\label{eq:passive_sys_a}
\begin{split}
da(t)
=&\;  \mathcal{A}a(t)dt+\mathcal{B}dB(t),
 \\
dB_{\mathrm{out}}(t) 
=&\;  \mathcal{C}a(t)dt+ \mathcal{D}dB(t),
\ \ t\geq t_0,
\end{split}
\end{equation}
where 
\begin{equation*}  \label{eq:passive_sys_ABCD}
\mathcal{A}=-i\Omega _{-}-\frac{1}{2}C_{-}^{\dagger }C_{-},\ \mathcal{B}%
=-C_{-}^{\dagger }S,\ \mathcal{C}=C_{-}, \   \mathcal{D} = S.
\end{equation*}
An equivalent way to characterize the structure of the passive linear quantum system \eqref{eq:passive_sys_a} is by the physical realizability conditions 
\begin{equation*}  \label{eq:passive_PR}
\mathcal{A}+\mathcal{A}^{\dagger }+\mathcal{B}\mathcal{B}^{\dagger }=0, ~\mathcal{B}=-\mathcal{C}^{\dagger }S.
\end{equation*}
Moreover, in the passive case,  $\Xi_{G^+}[s] \equiv 0 $ and 
 \begin{equation*}
 \Xi_{G^-}[s]  = S -C_-(sI+i\Omega _{-}+\frac{1}{2}C_{-}^{\dagger }C_{-})^{-1}C_-^\dag S.
 \end{equation*}
 Hence, if a linear system is passive, then its dynamics are completely characterized by its  annihilation operators.  Moreover, it can be easily verified that 
 \beqm
 \Xi_{G^-}[i\omega]^
 \dag  \Xi_{G^-}[i\omega] \equiv I_m, ~~~ \forall \omega\in \mathbb{R}.
 \eeqm
 Hence, an empty cavity does not change the amplitude of the input signal, but modifies its phase.

%%%%%%%%%%%%%%%%%%%%%%%%%%%%%%%%%%%%%
\bex[Re-visit Example \ref{ex:cavity}]\label{ex:cavity2}
For the cavity model in Example \ref{ex:cavity}, clearly $C^+=0$, $\Omega^+=0$ and $S=1$. In this case
 $\Xi_{G^+}[s] \equiv 0 $ and 
 \begin{equation*}
 \Xi_{G^-}[s]  = \frac{s+i\omega_d-\frac{\kappa}{2}}{s+i\omega_d+\frac{\kappa}{2}}.
 \end{equation*}
As  a result, the input-output relation in the frequency domain is 
\beqm
b_{{\rm out}}[s] =   \frac{s+i\omega_d-\frac{\kappa}{2}}{s+i\omega_d+\frac{\kappa}{2}} b_{{\mm[in]}}[s].
\eeqm
%\beq
%b_{{\rm out}}[i\omega] =   \frac{i\omega+i\omega_d-\frac{\kappa}{2}}{i\omega+i\omega_d+\frac{\kappa}{2}} b_{{\mm[in]}}[i\omega].
%\eeq
%Moreover, we have
%\beq
%a[s] = -\frac{\sqrt{\kappa}}{s+i\omega_d+\frac{\kappa}{2}} b_{{\mm[in]}}[s].
%\eeq
%\cite[ (7.28)]{WM08}
%\beq
%a_{{\rm out}}[-i\omega]  = - \frac{i\omega+i\omega_d-\frac{\kappa}{2}}{i\omega+i\omega_d+\frac{\kappa}{2}} a_{{\mm[in]}}[-i\omega].
%\eeq
%\beq
%a[-i\omega] = \frac{-\sqrt{\kappa}}{-i(\omega-\omega_d)+\frac{\kappa}{2}} b_{{\mm[in]}}[-i\omega].
%\eeq
\eex

Let the linear quantum system $G$ be initialized in the coherent state $\ket{\eta}$ and the input field be initialized in the vacuum state $\ket{0}$. Then the initial joint system-field state is $\rho_{0g}\triangleq \ket{\eta}\bra{\eta}
\otimes \ket{0}\bra{0}$ in the form of a density matrix.  Denote
\begin{equation*}\label{eq:rho_inf_g}
\rho_{\infty g} = \lim_{t\rightarrow\infty,t_{0}\rightarrow-\infty}U\left(
t,t_{0}\right)  \rho_{0g}U\left(t,t_{0}\right) ^{\ast}.
\end{equation*}
Here, $t_0\to -\infty$ indicates that the interaction starts in the remote past and $t\to\infty$ means that we are interested in the dynamics in the far future. In other words, we look at the steady-state dynamics. 
Define
\begin{equation}\label{eq:rho_field}
\rho_{\rm field,g}\triangleq  \langle \eta |\rho_{\infty g}| \eta \rangle.
\end{equation}
In other words, the system is traced off and we focus on the  steady-state  state of the output field.

The following result gives the response of a quantum linear system to a single-channel single-photon state.

%%%%%%%%%%%%%%%%%%%%%%%%%%%%%%%%%%%%%
\begin{theorem} \cite[Proposition 2]{ZJ13}   \label{prop:out-state-photon} 
Assume there is one input field which is in the single photon state $ \vert 1_\xi \rangle$. Also, assume that $G$ is Hurwitz stable and is initialized in a coherent state $\ket{\eta}$. Then the steady-state output field state for the linear quantum system $G$  is 
\begin{equation*}
\rho_{\rm out} = ( {\bf B}^\ast( \xi^-_{\rm out} ) - {\bf B}(\xi^+_{\rm out}) ) \rho_{\rm field,g} ( {\bf B}^\ast( \xi^-_{\rm out} ) - {\bf B}(\xi^+_{\rm out}) )^\ast ,
\label{eq:out-state-photon}
\end{equation*}
where
\begin{equation*}
\Delta ( \xi^-_{\rm out}[s] , \xi^+_{\rm out}[s] ) =  \Xi_{G}[s] \Delta ( \xi[s] , 0 ) ,
\end{equation*}
and $\rho_{\rm field, g}$, defined in  (\ref{eq:rho_field}), is the density operator for the output field with zero mean and covariance function
\begin{equation*} \label{eq:R_out}
R_{\rm out}[i\omega] =  \Xi_G[ i\omega] R_\mm[in][i\omega]  \Xi_G[i\omega]^\dag
\end{equation*}
with
\begin{equation*} \label{R_in_vacuum}
R_\mm[in][i\omega] = \left[
\begin{array}{ll}
 1 & 0 \\
  0 & 0
\end{array}
\right].
\end{equation*}
In particular, if the linear system $G$ is passive and initialized in the vacuum state, then  $\xi^+_{\rm out}[s] \equiv 0$ and $
R_{\rm out} [i\omega]  \equiv R_\mm[in][i\omega]$. 
In other words, the steady-state output is a single-photon state $\ket{1_{\xi^-_{\rm out}}}$.
\end{theorem}

%%%%%%%%%%%%%%%%%%%%%%%%%%%%%%%%%%%%%
\bex\label{ex:xi_out}
Let the optical cavity introduced in Example \ref{ex:cavity} be initialized in the vacuum state. Then, by Theorem \ref{prop:out-state-photon}, the steady-state output field state is also a single-photon state  $\ket{1_{\xi^-_{\rm out}}}$ with the pulse shape
\begin{equation*}\label{48}
 \xi^-_{\rm out}[i\omega]=\frac{i(\omega+\omega_d)-\frac{\kappa}{2}}{i(\omega+\omega_d)+\frac{\kappa}{2}}\xi[i\omega].
\end{equation*}
\eex

%%%%%%%%%%%%%%%%%%%%%%%%%%%%%%%%%%%%%
\bmrk
{\rm It has been shown in \cite{PZJ16} that the output field of a two-level atom initialized in the ground state and driven by a single-photon field $\ket{1_\xi}$ is also a single-photon state $\ket{1_{\xi^-_{\rm out}}}$. Thus, although the dynamics of a two-level atom is bilinear, see Example \ref{ex:atom}, in the single-photon input case it can be fully characterized by a linear systems theory.
}
\emrk

If the linear system $G$ is not passive, or is not initialized in the vacuum state, the steady-state output field state $\rho_{\rm out}$ in general is not a single-photon state; as can be seen in Theorem \ref{prop:out-state-photon}. This new type of states has been named ``photon-Gaussian'' states in \cite{ZJ13}. Moreover, it has been proved in  \cite{ZJ13} that the class of ``photon-Gaussian'' states is invariant under the steady-state action of a linear quantum system.  In what follows we present this result.

%%%%%%%%%%%%%%%%%%%%%%%%%%%%%%%%%%%%%
\begin{definition}\cite[Definition 1]{ZJ13} \label{def:F}
A state $\rho_{\xi, R}$ is said to be a \emph{photon-Gaussian} state if it belongs to the set
\begin{eqnarray*}
\mathcal{F} &\triangleq& \left\{\rho_{\xi, R} = \prod\limits_{k=1}^{m}\sum_{j=1}^{m}\left(B_j^\ast (\xi_{jk}^-) -B_j(\xi_{jk}^+) \right)\rho_R\left(\prod\limits_{k=1}^{m}\sum_{j=1}^{m}\left(B_j^\ast (\xi_{jk}^-) -B_j(\xi_{jk}^+) \right)\right)^\ast \right. 
\label{class_F}
\\
& & \ \ \ \   \left.  : \mathrm{function~}  \xi=\Delta(\xi^-, \xi^+) \mathrm{~and~ density~matrix~} \rho_R \mathrm{~satisfy ~} \mathrm{Tr}[\rho_{\xi, R}] = 1   \right\}. 
\nonumber
\end{eqnarray*}
\end{definition}

%%%%%%%%%%%%%%%%%%%%%%%%%%%%%%%%%%%%%
\begin{theorem}\cite[Theorem 5]{ZJ13}   \label{thm:main}
 Let  $\rho_{\xi_\mm[in], R_\mm[in]}  \in \mathcal{F}$ be a photon-Gaussian input state.  Also, assume that $G$ is Hurwitz stable and is initialized in a coherent state $\ket{\eta}$.   Then the linear quantum system $G$ produces in steady state a photon-Gaussian output state
$\rho_{\xi_{\rm out}, R_{\rm out}} \in \mathcal{F}$, where
\begin{eqnarray*}
 \xi_{\rm out}[s]  &=&  \Xi_{G}[s] \xi_\mm[in][s] ,  \label{eq:xi_out} \\
R_{\rm out}[i\omega] &=&  \Xi_G[ i\omega] R_\mm[in][i\omega]  \Xi_G[i\omega]^\dag  . \label{eq:R_out_gnr}
\end{eqnarray*}
\end{theorem}

Next, we present a result for the passive case, which is a special case of Theorem \ref{thm:main}.

Let the $k$th input channel be in a single photon state $\vert 1_{\mu_k} \rangle$, $k=1,\ldots, m$. Thus, the state of the $m$-channel input is given by the tensor product
\begin{equation}\label{eq:multichannel}
\vert \Psi_{\mu} \rangle  = \vert 1_{\mu_1} \rangle \otimes \cdots \otimes  \vert 1_{\mu_m} \rangle .
\end{equation}
Denote $\mu = [\mu_1 \ \ \cdots \ \ \mu_m]^\top$.

%%%%%%%%%%%%%%%%%%%%%%%%%%%%%%%%%%%%%
\begin{corollary}  \label{cor:passive}
Assume that the passive linear quantum system \eqref{eq:passive_sys_a} is Hurwitz stable,  initialized in  the vacuum state and driven by an $m$-photon input state  $\vert \Psi_{\mu} \rangle$.   The the steady-state output state is another  $m$-photon $\vert \Psi_{\nu} \rangle$    whose pulse   $\nu = [\nu_1 \ \ \cdots \ \ \nu_m]^\top$ is given by 
\beqm
\nu[i\omega] =   \Xi_{G^-}[i\omega]  \mu[i\omega].
\eeqm
\end{corollary}

Response of quantum linear systems to multi-photon states has been studied in \cite{Z14,Z17}.   In particular, the multi-photon versions of Corollary \ref{cor:passive} can be found in \cite[Corollary 11]{Z14} and \cite[Theorems 2-7 ]{Z17}. Response of quantum nonlinear systems to multi-photon states has been studied in \cite{milburn08,PZJ16,DZA19b}.

We end this section with a final remark.

%%%%%%%%%%%%%%%%%%%%%%%%%%%%%%%%%%%%%
\bmrk{\rm
To derive the output pulse shapes of a quantum linear system $G$ driven by continuous-mode single- or multiple-photon states,  it is assumed in \cite{ZJ13,Z14,Z17} that the system $G$ is Hurwitz stable. Actually,  if the system $G$ is passive,  the results also hold even if $G$ is marginally stable. The reason is the following. By \cite[Corollary 4.1]{ZGPG18},  a passive  quantum linear system  can  only have a $co$ subsystem which is  both controllable and observable, and a $\bar{c}\bar{o}$ subsystem which  is neither controllable nor observable. Actually, the  $\bar{c}\bar{o}$ subsystem is a closed system, and hence the modes related to the $\bar{c}\bar{o}$ subsystem will not affect system's input-output behavior. As the system is  passive, by \cite[Corollary 4.1]{ZGPG18}, the  $\bar{c}\bar{o}$ subsystem exactly corresponds to the subsystem whose poles are on the imaginary axis, and all the poles of the $co$ subsystem lie on the open left half of the complex plane, which means that the  $co$ subsystem is Hurwitz stable, and thus all the results in  \cite{ZJ13,Z14,Z17}  still hold.
}
\emrk

%%%%%%%%%%%%%%%%%%%%%%%%%%%%%%%%%%%%%
%%%%%%%%%%%%%%%%%%%%%%%%%%%%%%%%%%%%%
%%%%%%%%%%%%%%%%%%%%%%%%%%%%%%%%%%%%%
\section{Single-photon pulse shaping via coherent feedback}\label{sec:shaping}
In this section, based on the development in Section \ref{sec:linear_resp},  we demonstrate how a quantum linear coherent feedback network can be constructed to manipulate the temporal pulse shape of a single-photon input state.

\begin{figure}
\centering
\includegraphics[width=0.8\textwidth]{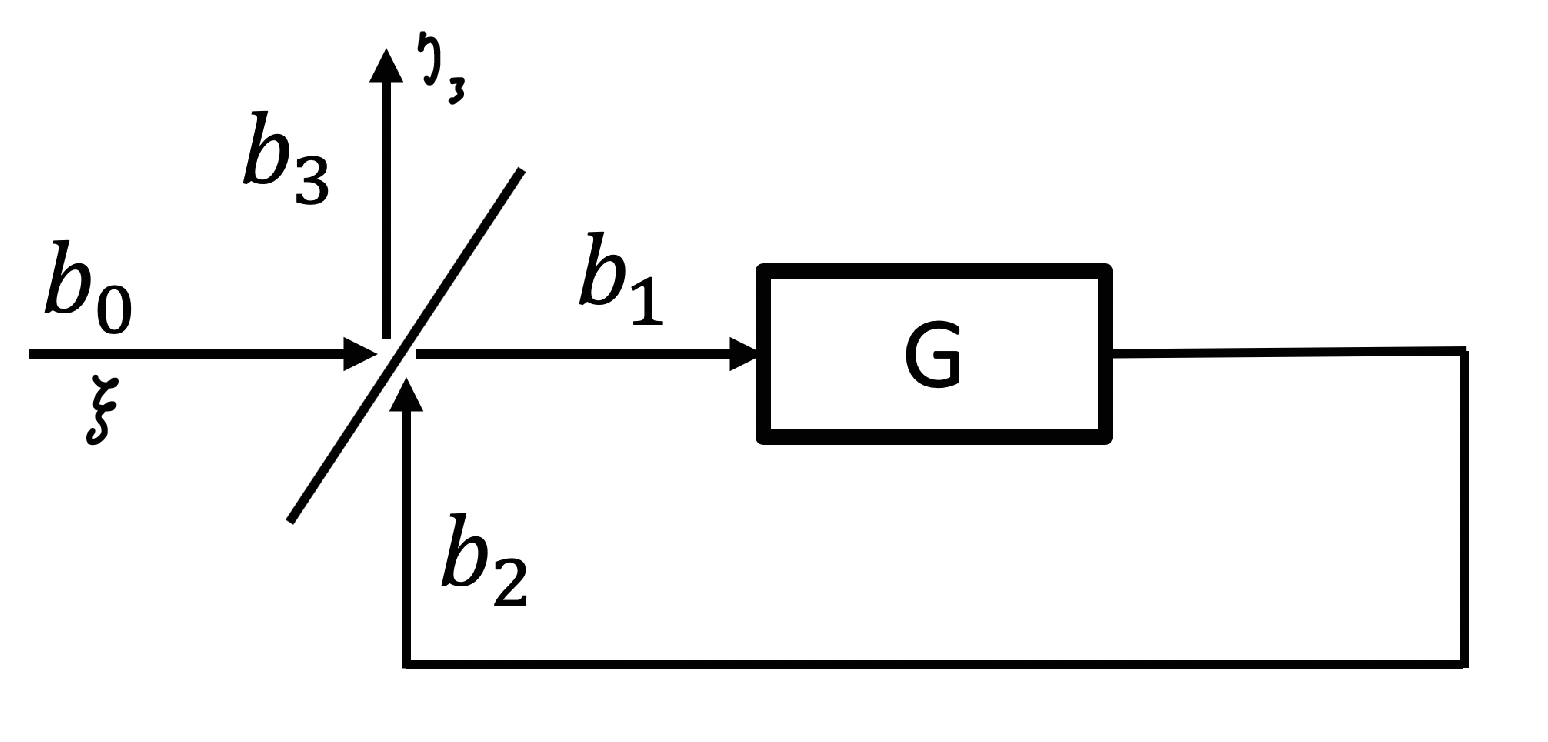}
\caption{Linear quantum coherent feedback network composed of an empty cavity and a beamsplitter. The input field $b_0$ is in the single photon state $\ket{1_\xi}$ and the output field $b_3$ is in the single-photon output state $\ket{1_{\eta_3}}$}
\label{fig_34}
\end{figure}

If a cavity, as given in Example \ref{ex:cavity}, is driven by a  single-photon state $\ket{1_\xi}$,  by Example \ref{ex:xi_out} the output pulse shape in the frequency domain is
\begin{equation}\label{48_feb10}
\eta_1[i\omega]= \frac{i(\omega+\omega_d)-\frac{\kappa}{2}}{i(\omega+\omega_d)+\frac{\kappa}{2}}\xi[i\omega].
\end{equation}

Now we put the cavity into a coherent feedback network closed by a beamsplitter, as shown in Fig.~\ref{fig_34}. (Here, the word ``coherent'' indicates that no measurement is involved in the feedback loop and thus all the signals remain quantum). Let the beamsplitter be
\begin{equation*}
S_{\rm BS}=\left[
      \begin{array}{cc}
        \sqrt{\gamma} & \sqrt{1-\gamma} \\
        -\sqrt{1-\gamma} & \sqrt{\gamma} \\
      \end{array}
    \right],  \ 0\leq\gamma\leq1.
\end{equation*}
Clearly, the beamsplitter $S_{\rm BS}$ is a special passive linear system  \eqref{eq:passive_sys_a} whose system parameters   are
\[
\mathcal{A}= \mathcal{B}  = \mathcal{C}=0, \   \mathcal{D} = S_{\rm BS}.
\]
Thus, the input-output relation for the beamsplitter $S_{\rm BS}$  is
\[\left[
\begin{array}{c}
b_3\\
b_1
\end{array}
\right] =S_{\rm BS} \left[
\begin{array}{c}
b_0\\
b_2
\end{array}
\right].
\]
Clearly, the feedback network from input $b_0$ to output $b_3$ in Fig.~\ref{fig_34} is still a quantum linear passive system that is driven by the single-photon state  $|1_{\xi}\rangle$ for the input field $b_0$. By the development in Section \ref{sec:linear_resp}, we can get the pulse shape for the output field $b_3$, which is
\begin{equation*}\label{52}
\eta_3[i\omega]=\frac{-\displaystyle\frac{1-\sqrt{\gamma}}{1+\sqrt{\gamma}}(\omega+\omega_d)i+\frac{\kappa}{2}}
{\displaystyle\frac{1-\sqrt{\gamma}}{1+\sqrt{\gamma}}(\omega+\omega_d)i+\frac{\kappa}{2}}\xi[i\omega].
\end{equation*}

\begin{figure}
\centering
\includegraphics[width=0.8\textwidth]{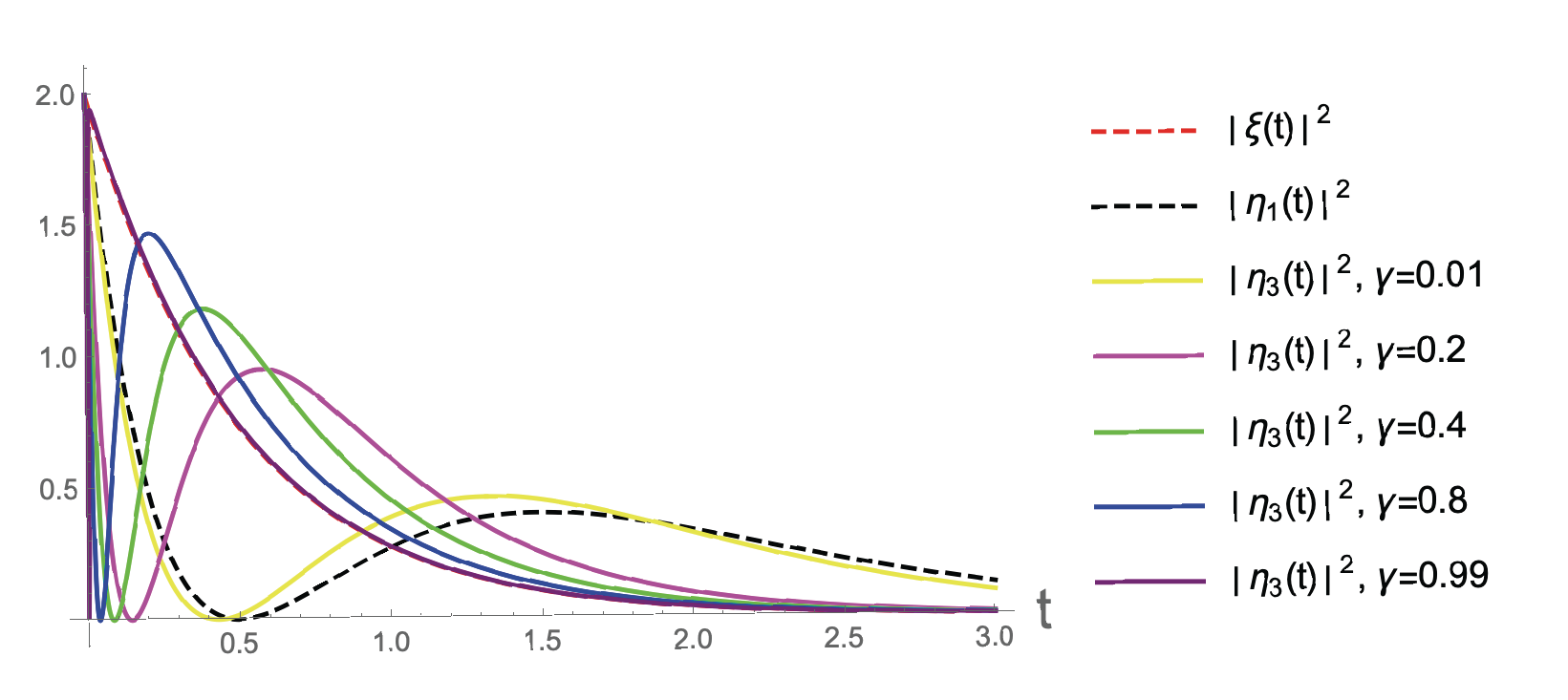}
\caption{$|\xi(t)|^2$ is the detection probability of the input single photon, $|\eta_1(t)|^2$ is  the detection probability of the output photon in the case of the cavity, $|\eta_3(t)|^2$ are the detection probabilities of the output photon in the linear coherent feedback network (Fig.~\ref{fig_34}) with various beamsplitter parameter $\gamma$.}
\label{fig_43}
\end{figure}

Fix $\beta = 2$ for the exponentially decaying single-photon state in  \eqref{31}, and  $\omega_d = 0$  and $\kappa =2$  for the optical cavity in  Example \ref{ex:cavity}.  The {\it temporal} pulse shapes $\xi(t)$, $\eta_1(t)$ and $\eta_3(t)$ are plotted in Fig.~\ref{fig_43}.

\begin{figure}
\centering
\includegraphics[width=0.8\textwidth]{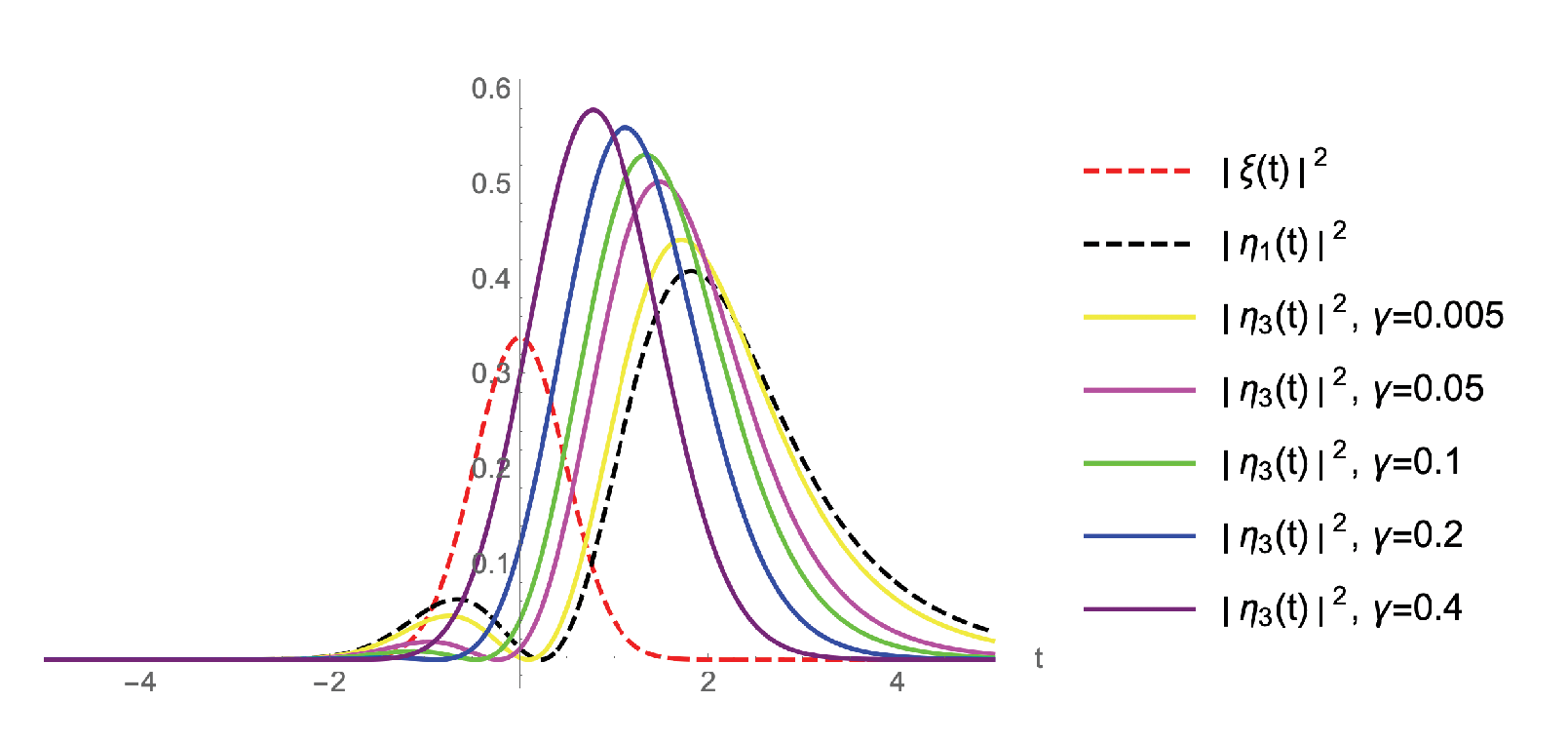}
\caption{The same as those in Fig.~\ref{fig_43}.}
\label{fig_gaussian}
\end{figure}

Fix $\tau = 0$ and $\Omega=1.46$ for a Gaussian single-photon state in  \eqref{51}, and $\omega_d = 0$ and $\kappa =1$ for the optical cavity.
When $\xi(t)$  is of a Gaussian pulse shape  (\ref{51})  for the optical cavity in  Example \ref{ex:cavity}.  The {\it temporal} pulse shapes $\xi(t)$, $\eta_1(t)$ and $\eta_3(t)$ are plotted in Fig.~\ref{fig_gaussian}.

%%%%%%%%%%%%%%%%%%%%%%%%%%%%%%%%%%%%%
%%%%%%%%%%%%%%%%%%%%%%%%%%%%%%%%%%%%%
%%%%%%%%%%%%%%%%%%%%%%%%%%%%%%%%%%%%%
\section{Single-photon filter and master equation}\label{sec:filtering}

\begin{figure}
\centering
\includegraphics[width=0.9\textwidth]{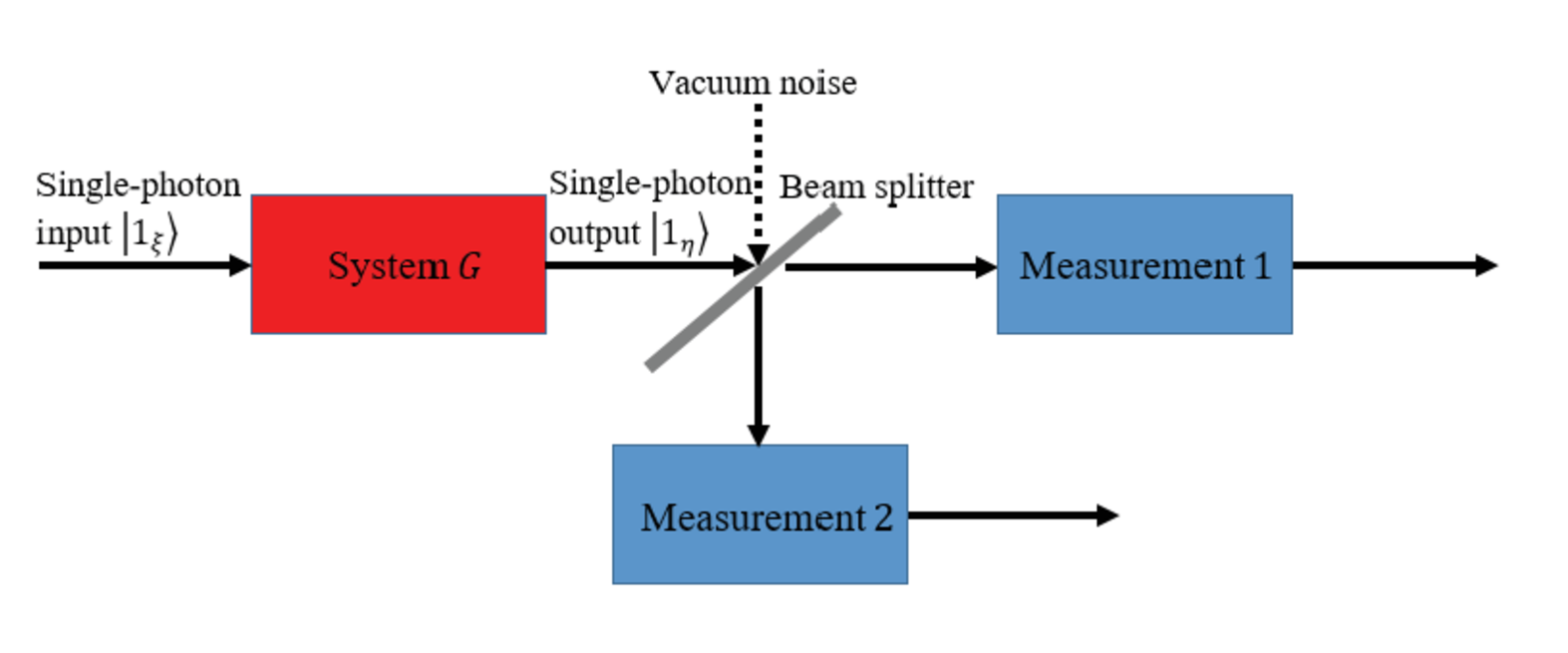}
\caption{Single-photon filtering. The system of interest $G$ is driven by a single-photon field state $\ket{1_\xi}$. The output single-photon state is denoted by $\ket{1_\eta}$. To account for the imperfectness of experiment \cite{SKH13}, a beamsplitter is added which mixes $\ket{1_\eta}$ with a vacuum noise. Both of the signals in the output ports of the beamsplitter are measured to improving the filtering effect.}
\label{fig_3}
\end{figure}

As discussed in Section \ref{sec:photon_state}, a single-photon light field has statistical properties. Therefore, it is natural to study the filtering problem of a quantum system driven by a single-photon field state.  Single-photon filters were first derived in \cite{GJN+12,GJN+12b}, and their multi-photon version was developed in \cite{SZX16,BC17,DZA19a}. In this section, we  focus on the single-photon case. The basic setup is given in Fig.~\ref{fig_3}.

The output field of an open quantum system can be continuously measured, see for example Subsection \ref{subsec:digital}; based on the measurement data a quantum filter can be built to estimate some quantity of the system. For example, we desire to know which state a two-level atom is in, the ground state $\ket{g}$ or  the excited state $\ket{e}$. Unfortunately, it is not realistic to measure the state of the atom directly. Instead, a light field may be impinged on the atom and from the scattered light we estimate the state of the atom.  Homodyne detection and photon-counting measurements are the two most commonly used measurement methods in quantum optical experiments. In this survey, we focus on Homodyne detection as discussed in Subsection \ref{subsec:digital}. In Fig.~\ref{fig_3}, $G$ is a  quantum system which is driven a single photon of pulse shape $\xi$.  
After interaction, the output field, represented by its integrated annihilation operator $B_{\rm out}$ and creation operator $B_{\rm out}^\ast$, is also in a single-photon state with pules shape $\eta$. Due to measurement imperfection (measurement inefficiency), the output field $\ket{1_\eta}$ may be contaminated \cite{SKH13,RR15}. This is usually mathematically modeled by mixing $\ket{1_\eta}$ with an additional quantum vacuum through a beam splitter, as shown in Fig.~\ref{fig_3}.  The beam splitter in Fig.~\ref{fig_3} is of a general form
\begin{equation} \label{S_b}
S_{\rm BS}=\left[
       \begin{array}{ll}
         s_{11} & s_{12} \\
         s_{21} & s_{22} \\
       \end{array}
     \right]
\end{equation}
where $s_{ij}\in \mathbb{C}$. As a result, there are two final output fields, which are
\[\left[
\begin{array}{c}
B_{1,\mathrm{out}}\\
B_{2,\mathrm{out}}
\end{array}
\right] =S_{\rm BS} \left[
\begin{array}{c}
B_{\rm out}\\
B_v
\end{array}
\right],
\]
where  $B_v$ is the integrated annihilation operator for the additional quantum noise channel. The quadratures of the  outputs are continuously measured by homodyne detectors, which are given by
\begin{equation}\label{filter:output}
Y_{1}(t)=B_{1,\mathrm{out}}(t)+B_{1,\mathrm{out}}^\ast(t),~~Y_{2}(t)=-i(B_{\rm 2, out}(t)-B_{\rm 2, out}^\ast(t)).
\end{equation}
In other words, the amplitude quadrature of the first output field is measured, while for the second output field the phase quadrature is monitored. $Y_{i}(t)$ ($i=1,2$) enjoy the self-non-demolition property
\begin{equation*}
[Y_i(t),Y_j(r)]=0,~~t_0\leq r\leq t, ~~ i,j=1,2,
\end{equation*}
and the  non-demolition property
\begin{equation*}
[X(t),Y_i(r)]=0,~~t_0\leq r\leq t, ~~ i=1,2,
\end{equation*}
where $t_0$ is the time when the system and field start interaction. The quantum conditional expectation is defined as
\begin{equation*}
\pi_t(X)\triangleq\mathbb{E}[j_t(X)|\mathcal{Y}_t],
\end{equation*}
where $\mathbb{E}$ denotes the expectation with respect to the initial joint system-field state,  $j_t(X)$ is given in \eqref{eq:X}, and the commutative von Neumann algebra $\mathcal{Y}_t$ is generated by the past measurement observations $\{Y_1(s), Y_2(s): t_0\leq s\leq t\}$. The {\it conditioned} system density operator $\rho(t)$ can be obtained by means of $\pi_t(X)=\mathrm{Tr}\left[\rho(t)^\dagger X\right]$.  It turns out that $\rho(t)$ is a solution to a system of stochastic differential equations, which is called the {\it quantum filter} in the quantum control community or {\it quantum trajectories} in the quantum optics community.  quantum filtering theory was pioneered by Belavkin in the early  1980s \cite{B80}. More developments can be found in \cite{belavkin1989nondemolition,PK98,vHSM05,bouten2007introduction,barchielli2009quantum,WM09,GJN+12b,RR15b,SZX16,DSC17,BC17,CKS17,jug18,DZA19a,GZP19,GZP20,D20} and references therein.

In the extreme case that $S_{\rm BS}$ is a 2--by-2 identity matrix,  the single-photon state $\ket{1_\eta}$ and the vacuum noise are not mixed and  $\ket{1_\eta}$ is directly measured by ``Measurement 1''. This is the case that the output of the two-level system $G$ is perfectly measured. In this scenario, a quantum filter constructed based on  ``Measurement 1'' is sufficient for  the estimation of conditioned system dynamics, as constructed in  \cite{GJN+12,GJN+12b}.  However, for a general beam splitter of the form (\ref{S_b}), the output of the two-level system $G$ is contaminated by vacuum noise, using two measurements may improve estimation efficiency, as investigated in \cite{DZA18}.

The single-photon filter for the set-up in Fig.~\ref{fig_3} is given by the following result.

%%%%%%%%%%%%%%%%%%%%%%%%%%%%%%%%%%%%%
\begin{theorem}\cite[Corollary 6.1]{DZA18}\label{corollary1}
Let the quantum system $G=(S,L,H)$ in Fig.~\ref{fig_3} be initialized in the state $\ket{\eta}$ and   driven by a single-photon input field $\ket{1_\xi}$. Assume the output fields  are under two homodyne detection measurements (\ref{filter:output}). Then the quantum filter in the Schr\"{o}dinger picture is given by
{\small
\begin{eqnarray}\label{qpsch}
\begin{aligned}
&d\rho^{11}(t)
\\
=&\left\{\mathcal{L}_G^\star\rho^{11}(t)+[S\rho^{01}(t),L^\dag]\xi(t)+[L,\rho^{10}(t)S^\dag]\xi^\ast(t)
+(S\rho^{00}(t)S^\dag-\rho^{00}(t))|\xi(t)|^2\right\}dt\\
&+\left[s_{11}^\ast\rho^{11}(t)L^\dag+s_{11}L\rho^{11}(t)+s_{11}^\ast\rho^{10}(t)S^\dag\xi^\ast(t)+
s_{11}S\rho^{01}(t)\xi(t)-\rho^{11}(t)k_1(t)\right]dW_1(t)\\
&+\left[is_{21}^\ast\rho^{11}(t)L^\dag-is_{21}L\rho^{11}(t)+is_{21}^\ast\rho^{10}(t)S^\dag\xi^\ast(t)-
is_{21}S\rho^{01}(t)\xi(t)-\rho^{11}(t)k_2(t)\right]dW_2(t),
\\
&d\rho^{10}(t)
\\
=&\left\{\mathcal{L}_G^\star\rho^{10}(t)+[S\rho^{00}(t),L^\dag]\xi(t)\right\}dt\\
&+\left[s_{11}^\ast\rho^{10}(t)L^\dag+s_{11}L\rho^{10}(t)+
s_{11}S\rho^{00}(t)\xi(t)-\rho^{10}(t)k_1(t)\right]dW_1(t)\\
&+\left[is_{21}^\ast\rho^{10}(t)L^\dag-is_{21}L\rho^{10}(t)-
is_{21}S\rho^{00}(t)\xi(t)-\rho^{10}(t)k_2(t)\right]dW_2(t),
\\
&d\rho^{00}(t)
\\
=&\mathcal{L}_G^\star\rho^{00}(t)dt+\left[s_{11}^\ast\rho^{00}(t)L^\dag+s_{11}L\rho^{00}(t)-\rho^{00}(t)k_1(t)\right]dW_1(t)\\
&+\left[is_{21}^\ast\rho^{00}(t)L^\dag-is_{21}L\rho^{00}(t)-\rho^{00}(t)k_2(t)\right]dW_2(t),\\
&\rho^{01}(t)=(\rho^{10}(t))^\dagger,
\end{aligned}
\end{eqnarray}
}
where $dW_j(t) = dY_j(t)-k_j(t)dt$ with
\begin{align*}
k_1(t)=&s_{11}\mathrm{Tr}[L\rho^{11}(t)]+s_{11}^\ast\mathrm{Tr}[L^\dag\rho^{11}(t)]
+s_{11}\mathrm{Tr}[S\rho^{01}(t)]\xi(t)
\\
&+s_{11}^\ast\mathrm{Tr}[S^\dag\rho^{10}(t)]\xi^\ast(t),
\\
k_2(t)=&-is_{21}\mathrm{Tr}[L\rho^{11}(t)]+is_{21}^\ast\mathrm{Tr}[L^\dag\rho^{11}(t)]
\\
&
-is_{21}\mathrm{Tr}[S\rho^{01}(t)]\xi(t)+is_{21}^\ast\mathrm{Tr}[S^\dag\rho^{10}(t)]\xi^\ast(t).
\end{align*}
 The initial conditions are $\rho^{11}(t_0)=\rho^{00}(t_0)=|\eta\rangle\langle\eta|$, $\rho^{10}(t_0)=\rho^{01}(t_0)=0$.
\end{theorem}

%%%%%%%%%%%%%%%%%%%%%%%%%%%%%%%%%%%%%
\bmrk{\rm
If the beamsplitter $S_{\rm BS}$ is a 2-by-2 identity matrix, the single-photon filter (\ref{qpsch}) in Theorem \ref{corollary1} reduces to
\begin{eqnarray}\label{qpsch_b}
\begin{aligned}
d\rho^{11}(t)
=&
\left\{\mathcal{L}_G^\star\rho^{11}(t)+[\rho^{01}(t),L^\dag]\xi(t)+[L,\rho^{10}(t)]\xi^\ast(t)\right\}dt
\\
&+\left[\rho^{11}(t)L^\dag+L\rho^{11}(t)+\rho^{10}(t)\xi^\ast(t)+
\rho^{01}(t)\xi(t)-\rho^{11}(t)k_1(t)\right]dW_1(t)
\\
d\rho^{10}(t)
=&
\left\{\mathcal{L}_G^\star\rho^{10}(t)+[\rho^{00}(t),L^\dag]\xi(t)\right\}dt\\
&+\left[\rho^{10}(t)L^\dag+L\rho^{10}(t)+\rho^{00}(t)\xi(t)-\rho^{10}(t)k_1(t)\right]dW_1(t),
\\
d\rho^{00}(t)
=&
\mathcal{L}_G^\star\rho^{00}(t)dt+\left[\rho^{00}(t)L^\dag+L\rho^{00}(t)-\rho^{00}(t)k_1(t)\right]dW_1(t),
\\
\rho^{01}(t)=&(\rho^{10}(t))^\dagger,
\end{aligned}
\end{eqnarray}
where $dW_1(t)$ and the initial conditions are the same as those in Theorem \ref{corollary1}, and 
\[
k_1(t)=\mathrm{Tr}[(L+L^\dag)\rho^{11}(t)]
+\mathrm{Tr}[\rho^{01}(t)]\xi(t)+\mathrm{Tr}[\rho^{10}(t)]\xi^\ast(t).
\]
The filter (\ref{qpsch_b}) is the quantum single-photon filter first proposed in \cite{GJN+12}.
}
\emrk

Quantum filters describe the joint system-field dynamics conditioned on measurement outputs; On the other hand, if the output field is traced out, we can get the master equation which describes the system dynamics. Master equations are regarded as unconditional system dynamics, see e.g., \cite{barchielli2009quantum,WM09,GJN+12}.  Setting $S=I$ and tracing over the noise terms (represented by $dW_1(t)$ and $dW_2(t)$) in the quantum filters \eqref{qpsch} and (\ref{qpsch_b}), we get the single-photon master equations in the Schr\"{o}dinger picture
\begin{eqnarray}\label{mesch}\begin{aligned}
\dot{\varrho}^{11}(t)=&\mathcal{L}^\star_G\varrho^{11}(t)+[\varrho^{01}(t),L^\dag]\xi(t)+[L,\varrho^{10}(t)]\xi^\ast(t),\\
\dot{\varrho}^{10}(t)=&\mathcal{L}^\star_G\varrho^{10}(t)+[\varrho^{00}(t),L^\dag]\xi(t),\\
\dot{\varrho}^{00}(t)=&\mathcal{L}^\star_G\varrho^{00}(t),\\
\varrho^{01}(t)=&(\varrho^{10}(t))^\dagger
\end{aligned}
\end{eqnarray}
with initial conditions $\varrho^{11}(t_0)=\varrho^{00}(t_0)=|\eta\rangle\langle\eta|$, $\varrho^{10}(t_0)=\varrho^{01}(t_0)=0$, where
\beqm
{\rm Tr}[\varrho^{jk}(t)^\dag X] = \bra{\eta\phi_j}j_t(X)\ket{\eta\phi_k}, \ \ j,k=0,1
\eeqm
with
\beqm
\ket{\phi_j}  = \left\{ 
\bey{ll}
\ket{0}, & j=0,\\
\ket{1_\xi}, &j=1.
\eey
\right.
\eeqm

The dynamics of a two-level atom driven by a single photon input field of Gaussian pulse shape has been studied intensively in the literature. In particular, when the photon has a Gaussian pulse shape (\ref{51}) with $\Omega=1.46\kappa$, where $\kappa$ is the decay rate of the two-level atom; see Example \ref{ex:atom}, it is shown that the maximal excitation probability is around $0.8$, see, e.g., \cite{SAL09}, \cite{RSF10}, \cite[Fig. 1]{WMS+11}, \cite[Fig. 8]{GJN+12}, and \cite[Fig. 2]{BCB+12}. Recently, the analytical expression of the pulse shape of the output single photon has been derived in \cite{PZJ16}, which is $\eta_1$ is  (\ref{48_feb10}). Assume the Gaussian pulse shape   in   (\ref{51})  has parameters $\tau=3$ and  $\Omega=1.46\kappa$.  It can be easily verified that $\int_{-\infty}^{{4}}\left(|\xi(\tau)|^2-|\eta_1(\tau)|^2\right)d\tau=0.8$. Interestingly, the excitation probability achieves its maximum 0.8 at time $t=4$ (the upper limit of the above integral). Therefore, the filtering result is consistent with that of the input-output response.

%%%%%%%%%%%%%%%%%%%%%%%%%%%%%%%%%%%%%
%%%%%%%%%%%%%%%%%%%%%%%%%%%%%%%%%%%%%
%%%%%%%%%%%%%%%%%%%%%%%%%%%%%%%%%%%%%
\section{Schr\"odinger cat states generation}\label{sec:cat}

Discussions in the previous sections are for continuous-mode {\it single-photon} states. In this section, we briefly discuss multi-photon states. In fact, a beamsplitter, which is a very simple linear quantum system but extremely widely used in optics, is able to entangle two input photons (one in each input port) such that two photons can coexist in a single output port. In other words, a 2-photon state is generated. A general theory of multi-photon processing by quantum linear systems has been developed in \cite{Z14,Z17}. In this section, we show how this theory can be applied to generate  Schr\"odinger cat states. 

Single-mode coherent states are defined in \eqref{eq:coherent state}. A Schr\"odinger  cat state is  a superposition state of two coherent states of opposite phase, 
$\ket{\beta}$ and $\ket{-\beta}$.  For example, the odd cat state is
\beqnm
  |\psi\rangle &\triangleq&
  {\cal N}_{-}(\ket{\beta}-\ket{-\beta})
                           ={\cal N}_{-}e^ \frac{-|\beta|^2}{2}\sum_{n=0}^{\infty}\frac{2\beta^{2n+1}}
                           {\sqrt{(2n+1)!}}|2n+1\rangle
\nonumber
\\
&=&
\sum_{n=0}^{\infty} \gamma_{\mm[cat],n}|2n+1\rangle ,                     
\label{equ1}
\eeqnm
where ${\cal N}_{-} = \ff{1}{\sqrt{2(1-e^{-2|\beta|^2})}}$ normalizes the odd cat state $|\psi\rangle$, and the amplitudes
\beqm
\gamma_{\mm[cat],n} \triangleq 
{\cal N}_{-}e^ \frac{-|\beta|^2}{2}\frac{2\beta^{2n+1}} {\sqrt{(2n+1)!}}.
\eeqm  

The bigger $|\beta|$ is, the larger the cat is.   Applications of  Schr\"odinger cat states in  quantum teleportation, quantum computation, and quantum metrology  can be found in e.g.,\cite{serafini2004minimum,BR08,neergaard2013quantum}. A scheme for generating Schr\"odinger  cat states is proposed in \cite{SKH13}.  In what follows, we use the linear systems theory developed in \cite{ZJ13,Z14,Z17} to derive the main equations in \cite{SKH13}.

A single-channel continuous-mode $\ell$-photon state $\ket{\psi_\ell} $ can be defined as
\begin{equation*}\label{state}
\ket{\psi_\ell}
\triangleq
\frac{1}{\sqrt{N_\ell}}\prod\limits_{k=1}^\ell {\bf B}^\ast(\xi_k)\ket{0} ,
\end{equation*}
where $N_\ell$ is the normalization coefficient. If $\xi_1(t)\equiv\cdots\equiv\xi_\ell(t)\equiv\xi(t)$, then this state is called \textit{continuous-mode}  Fock state which has been intensely studied, see e.g., \cite[(3)]{GEP+98}; \cite[(13)]{BCB+12}. If we forget the pulse shapes, we can use
\beqm
\ket{n} =\frac{1}{\sqrt{n!}} ({\bf B}^\ast(\xi))^n \ket{0},  \ \ \|\xi\|_2=1
\eeqm
to denote an $n$-photon Fock state.  In this manner, an $m$-channel multi-photon tensor product state is of the form
\beqm
\ket{\Psi} = \ket{n_1}\otimes\ket{n_2} \otimes \cdots \otimes  \ket{n_m},
\eeqm
where the $j$th channel contains $n_i$ photons.

Consider a beamsplitter of the form
\beq \la{eq:BS}
S=\left[ 
\begin{array}{cc}
T & -R \\ 
R & T%
\end{array}%
\right] ,\ \ \ (R,T \in \mathbb{R},~~R^{2}+T^{2}=1).
\eeq
By  \cite[Corollary 11 or Example 3]{Z14}, it can be derived that the beamsplitter \eqref{eq:BS} maps a product state $ \ket{n_1}\otimes\ket{n_2}$ to an entangled output state
\beqn 
&&\ket{\Psi_\mm[out]} 
\label{eq:key_a}
\\
&=&
\ff{1}{\sqrt{n_1! n_2!}} \sum_{j=0}^{n_1}\sum_{k=0}^{n_2} \binom{n_1}{j}\binom{n_2}{k} 
\nonumber
\\
&& 
(-1)^k T^{n_2+j-k} R^{n_1-j+k}\sqrt{(j+k)! (n_1+n_2-j-k)!}\ket{j+k}\ket{n_1+n_2-j-k}. 
\nonumber
\eeqn
An ideal single-mode squeezed vacuum state can be
expressed as 
\begin{equation}
\hat{S}(\eta )\ket{0} =\frac{1}{\sqrt{\cosh \eta }}\sum_{n=0}^{\infty }\alpha _{2n}\ket{2n} ,
\label{eq:aug10_squeezed_vacuum}
\end{equation}%
where%
\begin{equation*}
\alpha _{2n}=\frac{\sqrt{(2n)!}}{2^{n}n!}(-e^{i\phi})^n\tanh
^{n}\eta
  \label{eq:aug10_alpha_2n}
\end{equation*}
with $\eta\in\mathbb{R}$ being the squeezing ratio. In this article, we choose the squeezing angle $\phi=\pi$ for simplicity.  Then $\alpha_{2n}\in \mathbb{R}$. More discussions of squeezed states can be found in \cite[Chapter 5]{RL00}, \cite[Chapter 10]{GZ00},  \cite{AL15}, \cite{FT20},   \cite{ZWD+20}, and \cite{ATF+21}.    According to  \eqref{eq:key_a},  the beamsplitter \eqref{eq:BS} maps $\ket{\ell}\otimes \hat{S}(\eta )\ket{0}$ to 
\beqn
\ket{\Psi _\mm[out]}
&=& 
\sum_{n=0}^{\infty }\frac{\alpha _{2n}}{\sqrt{\ell!(2n)!}}
\sum_{i=0}^{\ell}\sum_{j=0}^{2n}\binom{\ell }{i}\binom{2n}{2n-j}
\label{eq:aug10_2}
\\
&& \sqrt{(\ell +j-i)!(2n+i-j)!}(-1)^{j}T^{2n+\ell -i-j}R^{i+j}|\ell +j-i\rangle \otimes |2n+i-j\rangle .
\nonumber
\eeqn
Let $\ell=0$, i.e., the first input channel is in the vacuum state. In this case,  $\ket{\Psi _\mm[out]}$ in \eqref{eq:aug10_2} in a density matrix form is
\beqn
&&\hat{\rho} = \ket{\Psi _\mm[out]} \bra{\Psi _\mm[out]}
\label{eq:mar3_rho}
 \\
&=& 
\sum_{n,b=0}^{\infty }\alpha _{2n} \alpha^\ast _{2b} 
\sum_{j_1=0}^{2n} \sum_{j_2=0}^{2b} \sqrt{\binom{2n}{2n-j_1}\binom{2b}{2b-j_2}}
(-1)^{j_1+j_2}T^{2(n+b)-(j_1+j_2)}R^{j_1+j_2}
\nonumber
\\
&& \ket{j_1}\ket{2n-j_1}   \bra{2b-j_2} \bra{j_2}.
\nonumber
\eeqn
If $k$ photons are subtracted in the {\bf first} output channel, then by  \eqref{eq:mar3_rho}  the unnormalized conditional state in the second output channel is 
\beqnm 
&&\mathrm{Tr}_1\{\hat{\rho}\}
\\ 
&=&  \sum_{k=0}^\infty  \bra{k}\hat{\rho} \ket{k}
\label{eq:feb10_h2}
\\
&=&
\sum_{n,b=0}^{\infty }\alpha _{2n} \alpha^\ast _{2b} \sum_{k=0}^{\min\{2n,2b\}}
\sqrt{\binom{2n}{2n-k}\binom{2b}{2b-k}}
T_1^{2(n+b-k)}R_1^{2k}\ket{2n-k}   \bra{2b-k}, 
\nonumber
\eeqnm
which is $\reallywidehat{\rho}_{t_1}$ in  \cite[ (5)]{SKH13}.  Actually, $\reallywidehat{\rho}_{t_1}$ is an {\it impure} squeezed vacuum state; which accounts for experimental imperfection when an ideal squeezed vacuum state \eqref{eq:aug10_squeezed_vacuum} is used to generate a Schr\"odinger cat state; see the red box in  \cite[Fig. 1]{SKH13}. Next, we derive the output density matrix which is the output of the green box in \cite[Fig. 1]{SKH13}. Define 
\beqnm
\rho_\mathrm{in,2}
&=&
\ket{\ell}\bra{\ell} \otimes  \hat{\rho}_{t_1}
\nonumber
\\
&=& 
\sum_{n,b=0}^{\infty }\alpha _{2n} \alpha^\ast _{2b} \sum_{k=0}^{\min\{2n,2b\}}
\sqrt{\binom{2n}{2n-k}\binom{2b}{2b-k}}
T_1^{2(n+b-k)}R_1^{2k}
\\
&&
\ket{\ell}\ket{2n-k}   \bra{2b-k} \bra{\ell}.
\nonumber
\eeqnm
In particular, if $\ell=0$, then $\rho_\mathrm{in,2}$ is the input to the beamsplitter in the green box in \cite[Fig. 1]{SKH13}. According to  \eqref{eq:key_a}, this beamsplitter maps the pure  state $\ket{\ell}\otimes \ket{2n-k}$ to
\beqnm 
&&\ket{\Psi_{2n-k,\ell}} 
\\
&=&
\ff{1}{\sqrt{\ell! (2n-k)!}} \sum_{j=0}^{\ell}\sum_{i=0}^{2n-k} \binom{\ell}{j}\binom{2n-k}{i} (-1)^i
\label{eq:key}
\\
&&R_2^{\ell-j+i}  T_2^{2n-k+j-i} \sqrt{(j+i)! (\ell+2n-k-j-i)!}\ket{j+i}\ket{\ell+2n-k-j-i}.
\nonumber 
\eeqnm
Consequently, the beamsplitter maps the input state  $\rho_\mathrm{in,2}$ to an output state of the form
\beqnm\label{eq:Feb11_a1}
&&\rho_\mathrm{out,2}
\\
 &=&
 \sum_{n,b=0}^{\infty }\alpha _{2n} \alpha^\ast _{2b} \sum_{k=0}^{\min\{2n,2b\}}
\sqrt{\binom{2n}{2n-k}\binom{2b}{2b-k}}
T_1^{2(n+b-k)}R_1^{2k} \ket{\Psi_{2n-k,\ell}}  \bra{\Psi_{2b-k,\ell}} .
\eeqnm
In particular, let $\ell=0$. If we measure the first output channel and detect $m$ photons, then the unnormalized reduced  density matrix is 
\beqn
&&\braket{m|\rho_\mathrm{out,2}|m}
\nonumber
\\
&=&
\sum_{n,b=0}^{\infty }\alpha _{2n} \alpha^\ast _{2b} \sum_{k=0}^{\min\{2n,2b\}-m}
\sqrt{\binom{2n}{2n-k}\binom{2b}{2b-k}}
T_1^{2(n+b-k)}R_1^{2k} 
\nonumber
\\
&&
\ff{1}{\sqrt{(2n-k)!(2b-k)!}} \binom{2b-k}{m}  \binom{2n-k}{m} R_2^{2m} T_2^{2(n+b-k-m)}  m!
\nonumber
\\ 
&&\sqrt{(2b-k-m)! (2n-k-m)! } \ket{2n-k-m}\bra{2b-k-m}
\nonumber
\\
&=&
\sum_{n,b=0}^{\infty }\alpha _{2n} \alpha^\ast _{2b} \sum_{k=0}^{\min\{2n,2b\}-m}
\sqrt{\binom{2n}{2n-k}\binom{2b}{2b-k}}
T_1^{2(n+b-k)}R_1^{2k} 
\nonumber
\\
&&
\sqrt{ \binom{2b-k}{m}  \binom{2n-k}{m}} R_2^{2m} T_2^{2(n+b-k-m)}   \ket{2n-k-m}\bra{2b-k-m}
\nonumber
\\
&=&
\sum_{n,b=0}^{\infty } \sum_{k=0}^{\min\{2n,2b\}-m} \frac{\alpha _{2n} \alpha^\ast _{2b}   R_1^{2k} R_2^{2m} (T_1 T_2)^{2(n+b-k)} T_2^{-2m} }{k! \;m!}
\nonumber
\\
&&
\sqrt{\frac{(2n)! (2b)!}{(2n-k-m)! (2b-k-m)!}}   \ket{2n-k-m}\bra{2b-k-m}.
\label{eq:key4}
\eeqn

%%%%%%%%%%%%%%%%%%%%%%%%%%%%%%%%%%%
\bmrk{\rm
Let $t_k = T_k^2$ and $r_k = R_k^2$, $k=1,2$. $\braket{m|\rho_\mathrm{out,2}|m}$ in \eqref{eq:key4} becomes   \cite[ (11)]{SKH13}, whose normalized version is the output density matrix of the green box in \cite[Fig. 1]{SKH13}.  (The term $T_2^{-2m}$ is missing in \cite[Fig. 1]{SKH13}, which is a typo.)
}
\emrk

%%%%%%%%%%%%%%%%%%%%%%%%%%%%%%%%%%%
\bmrk{\rm
Many schemes like those proposed in \cite{RGM+03,OJT+07,GGC+10,HLR+15,EBK+15,SUP+17,OJ18,ENP19,MPK+19,TYA+21}   that generate Schr\"odinger cat states can  be described uniformly in the mathematical framework similar to that discussed above. 
}
\emrk

%%%%%%%%%%%%%%%%%%%%%%%%%%%%%%%%%%%%%
%%%%%%%%%%%%%%%%%%%%%%%%%%%%%%%%%%%%%
%%%%%%%%%%%%%%%%%%%%%%%%%%%%%%%%%%%%%
\section{Concluding remarks}\label{sec:Con}

In this section, we discuss two possible future research directions.

%In Section \ref{sec:shaping}, we have discussed the problem of single-photon pulse shaping by using a very simple example, see Fig.~\ref{fig_34}, where the system $G$ is an optical cavity. Clearly, a passive quantum linear controller $K$ can be added into the network in Fig.~\ref{fig_34}; see Fig.~\ref{fig_34c}. If both the system $G$ and the controller $K$ are linear time invariant, that is, all their  parameters $\Omega_-$ and $C_-$ as discussed in Section  \ref{sec:linear_resp} are constant matrices, then the overall system from $b_0$ to $b_3$ is still a passive quantum linear system with constant system matrices. In this case, if the network is driven by a single photon, the  output channel $b_3$ contains a single photon whose temporal pulse shape can be derived by Theorem \ref{thm:main}. Clearly, the output pulse shape is a function of the physical parameters of the passive quantum controller $K$.  Thus, adding a passive quantum linear controller  may increase flexibility of the pulse shaping of the input single photon.  There may be one of the future research directions.
%
\begin{figure}
\centering
\includegraphics[width=0.6\textwidth]{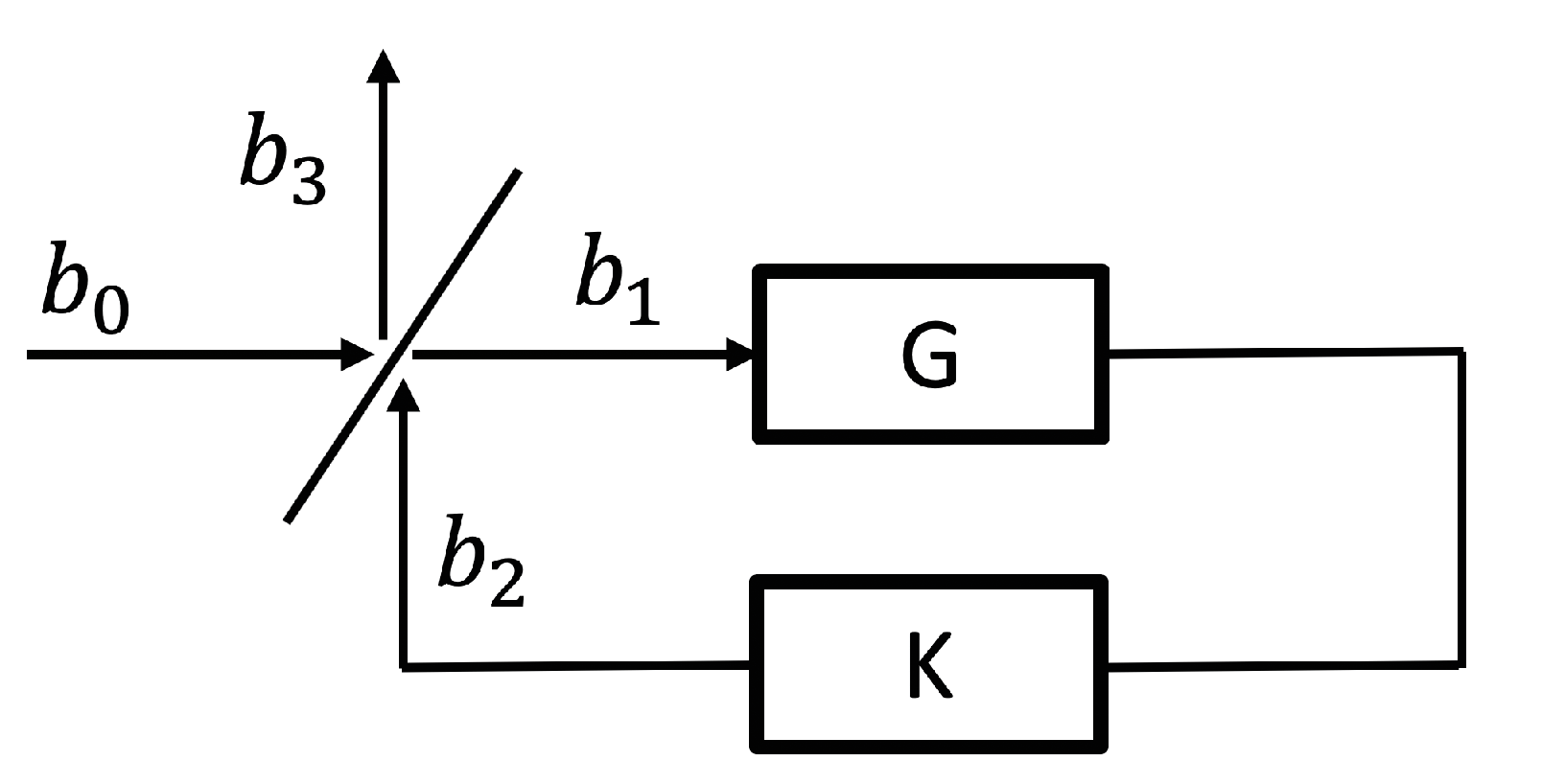}
\caption{Linear quantum feedback network}
\label{fig_34c}
\end{figure}

In \cite{milburn08}, Milburn investigated the response of an optical cavity to a continuous-mode single photon, where  frequency modulation applied to the cavity is used to engineer the temporal output pulse shape. As  frequency modulation involves a time-varying function, the transfer function approach in \cite{ZJ13} is not directly applicable. However, it appears that the general procedure outlined in the proof of Proposition 2 and Theorem 5 in \cite{ZJ13} can be generalized to the time-varying, yet still linear, case. For example, if the quantum linear passive controller $K$ in Fig.~\ref{fig_34c}  is allowed to be time-varying,  can the output single-photon pulse shape be engineered satisfactorily? This may be a future research direction.

In \cite{DCZ20}, the authors discussed how to use a single-photon input state to excite one of the two double quantum dot (DQD) qubits which are in a coherent feedback network; see Fig.~1 in \cite{DCZ20}.  A master equation of the form \eqref{mesch} is used to describe the reduced dynamics of the system. By tracing over the cavity and the 2nd DQD qubit, the reduced density operator for the target DQD qubit is given by
\begin{equation}\label{Oct30}
\rho_{\mathrm{DQD}_1}(t)=\langle g_20|\rho^{11}(t)|g_20\rangle+\langle g_21|\rho^{11}(t)|g_21\rangle+\langle e_20|\rho^{11}(t)|e_20\rangle+\langle e_21|\rho^{11}(t)|e_21\rangle,
\end{equation}
which is \cite[ (45)]{DCZ20}.
 Let the initial state of the two DQD qubits be the ground states $\ket{g_1}$ and  $\ket{g_2}$ and the cavity be initially empty.  The purpose of control is to flip the first DQD qubit to its excited state $\ket{e_1}$ at some time instance $T$, by means of designing a single-photon pulse; in other words, the single-photon pulse $\xi(t)$ is used as the control input. As $\xi$ is the pulse shape of the single photon, it is in the space $L^2(\mathbb{R},\mathbb{C})$ with additional constraint that is $L^2$ norm $\|\xi\|=1$. The physical interpretation of this constraint can be found in Section 2.2 in  \cite{DCZ20}. The state of the first DQD qubit evolves according to the master equation \eqref{Oct30}. As $\rho_\mathrm{DQD_1}$ in \eqref{Oct30} is the reduced density matrix of the first DQD qubit, thus it is a $2\times 2$ Hermitian matrix. For the cost function, we may use the Frobenius norm $\|\rho_{\mathrm{DQD}_1}(T)- \ket{e_1}\bra{e_1}\|^2_\mathrm{F}$. That is,  we desire to steer the state of the first DQD qubit as close as possible to $\ket{e_1}$ at a given time instance $T$.  Moreover,  as discussed in  \cite{BB20}, to get a sparse signal which is often desirable in experiments,  we may also add an $L^1$  term $\beta |\xi|_1$ in the cost function. In summary, to fit into (4.1) in  \cite{BB20},  we may choose the following  cost function
\begin{equation*}\label{eq1:dec11}
\|\rho_{\mathrm{DQD}_1}(T)- \ket{e_1}\bra{e_1}\|^2_\mathrm{F} + \beta \int_0^T |\xi(t)|_1 dt.
\end{equation*}
As $\xi(t)$ is the pulse shape of the single-photon state to be designed, quantum physics demands the $L^2$ norm $\|\xi\|_2 \triangleq \sqrt{\int_0^T |\xi(t)|^2 dt}=1$. Thus, in the above formulation of the optimal control problem, the to-be-designed pulse shape is in the set $\mathcal{C}=\{ \xi:\|\xi\|_2=1 \}$. Clearly, this constraint makes the admissible set of $\xi$ non-convex. In order to use quantum optimal control methods as those in   \cite{CB16,BCS17,BB20},  we have to remove this constraint.  One possible way is the following: Firstly, we ignore the constraint imposed by $\mathcal{C}$ and solve the optimization problem by means of quantum optimal control methods such as those in  \cite{CB16,BCS17,BB20}. Assume the obtained solution is $\xi(t)$ over the time interval $[0,T]$.  Then let $\gamma = \|\xi\|_2$. If  $\gamma=1$, it is perfect, nothing else is needed. Otherwise, define $\eta(t)\triangleq \gamma \xi\left(\frac{t}{\gamma }\right)$ over the time interval $[0,\gamma T]$. Then $\|\eta\|=1$,  we use $\eta$, a scaled version of $\xi$, instead of $\xi$ as the pulse shape to designed. Clearly, $\eta$ accomplishes the job at the terminal time $\gamma T$.  Solving this optimal control problem may be another future research direction.

\bibliographystyle{ieeetr}
\bibliography{/Users/zhang/Dropbox/gzhang}

\begin{thebibliography}{100}

\bibitem{LHA+01}
A.~I. Lvovsky, H.~Hansen, T.~Aichele, O.~Benson, J.~Mlynek, and S.~Schiller,
  ``Quantum state reconstruction of the single-photon {F}ock state,'' {\em
  Physical Review Letters}, vol.~87, no.~5, p.~050402, 2001.

\bibitem{YKS+02}
Z.~Yuan, B.~E. Kardynal, R.~M. Stevenson, A.~J. Shields, C.~J. Lobo, K.~Cooper,
  N.~S. Beattie, D.~A. Ritchie, and M.~Pepper, ``Electrically driven
  single-photon source,'' {\em Science}, vol.~295, no.~5552, pp.~102--105,
  2002.

\bibitem{MBB+04}
J.~McKeever, A.~Boca, A.~D. Boozer, R.~Miller, J.~R. Buck, A.~Kuzmich, and
  H.~J. Kimble, ``Deterministic generation of single photons from one atom
  trapped in a cavity,'' {\em Science}, vol.~303, no.~5666, pp.~1992--1994,
  2004.

\bibitem{HSG+07}
A.~A. Houck, D.~Schuster, J.~Gambetta, J.~Schreier, B.~Johnson, J.~Chow,
  L.~Frunzio, J.~Majer, M.~Devoret, S.~Girvin, {\em et~al.}, ``Generating
  single microwave photons in a circuit,'' {\em Nature}, vol.~449, no.~7160,
  pp.~328--331, 2007.

\bibitem{OFV09}
J.~L. O'{B}rien, A.~Furusawa, and J.~Vu{\v{c}}kovi{\'c}, ``Photonic quantum
  technologies,'' {\em Nature Photonics}, vol.~3, no.~12, pp.~687--695, 2009.

\bibitem{BC09}
G.~Buller and R.~Collins, ``Single-photon generation and detection,'' {\em
  Measurement Science and Technology}, vol.~21, no.~1, p.~012002, 2009.

\bibitem{LR09}
A.~I. Lvovsky and M.~G. Raymer, ``Continuous-variable optical quantum-state
  tomography,'' {\em Reviews of modern physics}, vol.~81, no.~1, p.~299, 2009.

\bibitem{SFY10}
C.~Santori, D.~Fattal, and Y.~Yamamoto, {\em Single-photon devices and
  applications}.
\newblock John Wiley \& Sons, 2010.

\bibitem{BRV12}
S.~Buckley, K.~Rivoire, and J.~Vu{\v{c}}kovi{\'c}, ``Engineered quantum dot
  single-photon sources,'' {\em Reports on Progress in Physics}, vol.~75,
  no.~12, p.~126503, 2012.

\bibitem{PHC+14}
M.~Pechal, L.~Huthmacher, C.~Eichler, S.~Zeytino{\u{g}}lu, A.~Abdumalikov~Jr,
  S.~Berger, A.~Wallraff, and S.~Filipp, ``Microwave-controlled generation of
  shaped single photons in circuit quantum electrodynamics,'' {\em Physical
  Review X}, vol.~4, no.~4, p.~041010, 2014.

\bibitem{LMS}
P.~Lodahl, S.~Mahmoodian, and S.~Stobbe, ``Interfacing single photons and
  single quantum dots with photonic nanostructures,'' {\em Reviews of Modern
  Physics}, vol.~87, no.~2, p.~347, 2015.

\bibitem{RR15}
A.~Reiserer and G.~Rempe, ``Cavity-based quantum networks with single atoms and
  optical photons,'' {\em Reviews of Modern Physics}, vol.~87, no.~4, p.~1379,
  2015.

\bibitem{NJY16}
H.~I. Nurdin, M.~R. James, and N.~Yamamoto, ``Perfect single device absorber of
  arbitrary traveling single photon fields with a tunable coupling parameter:
  {A QSDE} approach,'' in {\em 2016 IEEE 55th Conference on Decision and
  Control (CDC)}, pp.~2513--2518, IEEE, 2016.

\bibitem{OOM+16}
H.~Ogawa, H.~Ohdan, K.~Miyata, M.~Taguchi, K.~Makino, H.~Yonezawa, J.-i.
  Yoshikawa, and A.~Furusawa, ``Real-time quadrature measurement of a
  single-photon wave packet with continuous temporal-mode matching,'' {\em
  Phys. Rev. Lett.}, vol.~116, p.~233602, Jun 2016.

\bibitem{Peng2016}
Z.~H. Peng, S.~E.~D. Graaf, J.~S. Tsai, and O.~V. Astafiev, ``Tuneable
  on-demand single-photon source,'' {\em Nature Communications}, vol.~7,
  no.~12588, 2016.

\bibitem{GKM+17}
X.~Gu, A.~F. Kockum, A.~Miranowicz, Y.-x. Liu, and F.~Nori, ``Microwave
  photonics with superconducting quantum circuits,'' {\em Physics Reports},
  vol.~718, pp.~1--102, 2017.

\bibitem{DTK+18}
A.~O. Davis, V.~Thiel, M.~Karpi{\'n}ski, and B.~J. Smith, ``Measuring the
  single-photon temporal-spectral wave function,'' {\em Physical review
  letters}, vol.~121, no.~8, p.~083602, 2018.

\bibitem{WQD+19}
H.~Wang, J.~Qin, X.~Ding, M.-C. Chen, S.~Chen, X.~You, Y.-M. He, X.~Jiang,
  L.~You, Z.~Wang, {\em et~al.}, ``Boson sampling with 20 input photons and a
  60-mode interferometer in a $10^{14}$-dimensional {H}ilbert space,'' {\em
  Physical review letters}, vol.~123, no.~25, p.~250503, 2019.

\bibitem{TOS+19}
K.~Takase, M.~Okada, T.~Serikawa, S.~Takeda, J.-i. Yoshikawa, and A.~Furusawa,
  ``Complete temporal mode characterization of non-{G}aussian states by a dual
  homodyne measurement,'' {\em Physical Review A}, vol.~99, no.~3, p.~033832,
  2019.

\bibitem{SAL09}
M.~Stobi{\'n}ska, G.~Alber, and G.~Leuchs, ``Perfect excitation of a matter
  qubit by a single photon in free space,'' {\em EPL (Europhysics Letters)},
  vol.~86, no.~1, p.~14007, 2009.

\bibitem{WMS+11}
Y.~Wang, J.~Min{\'a}{\v{r}}, L.~Sheridan, and V.~Scarani, ``Efficient
  excitation of a two-level atom by a single photon in a propagating mode,''
  {\em Physical Review A}, vol.~83, no.~6, p.~063842, 2011.

\bibitem{PZJ16}
Y.~Pan, G.~Zhang, and M.~R. James, ``Analysis and control of quantum
  finite-level systems driven by single-photon input states,'' {\em
  Automatica}, vol.~69, pp.~18--23, 2016.

\bibitem{RSF10}
E.~Rephaeli, J.-T. Shen, and S.~Fan, ``Full inversion of a two-level atom with
  a single-photon pulse in one-dimensional geometries,'' {\em Physical Review
  A}, vol.~82, no.~3, p.~033804, 2010.

\bibitem{GJN+12}
J.~E. Gough, M.~R. James, H.~I. Nurdin, and J.~Combes, ``Quantum filtering for
  systems driven by fields in single-photon states or superposition of coherent
  states,'' {\em Physical Review A}, vol.~86, no.~4, p.~043819, 2012.

\bibitem{BCB+12}
B.~Q. Baragiola, R.~L. Cook, A.~M. Bra{\'n}czyk, and J.~Combes, ``N-photon wave
  packets interacting with an arbitrary quantum system,'' {\em Physical Review
  A}, vol.~86, no.~1, p.~013811, 2012.

\bibitem{SZX16}
H.~Song, G.~Zhang, and Z.~Xi, ``Continuous-mode multiphoton filtering,'' {\em
  SIAM Journal on Control and Optimization}, vol.~54, no.~3, pp.~1602--1632,
  2016.

\bibitem{DZA19a}
Z.~Dong, G.~Zhang, and N.~H. Amini, ``Quantum filtering for a two-level atom
  driven by two counter-propagating photons,'' {\em Quantum Information
  Processing}, vol.~18, no.~5, p.~136, 2019.

\bibitem{DZA19b}
Z.~Dong, G.~Zhang, and N.~H. Amini, ``On the response of a two-level system to
  two-photon inputs,'' {\em SIAM Journal on Control and Optimization}, vol.~57,
  no.~5, pp.~3445--3470, 2019.

\bibitem{GZ00}
C.~Gardiner and P.~Zoller, {\em Quantum Noise}.
\newblock Springer, 2004.

\bibitem{BP02}
E.~B. Davies, {\em Quantum Theory of Open Systems}.
\newblock Academic Press London, 1976.

\bibitem{WM09}
H.~M. Wiseman and G.~J. Milburn, {\em Quantum Measurement and Control}.
\newblock Cambridge university press, 2009.

\bibitem{GJ09}
J.~Gough and M.~R. James, ``The series product and its application to quantum
  feedforward and feedback networks,'' {\em IEEE Transactions on Automatic
  Control}, vol.~54, no.~11, pp.~2530--2544, 2009.

\bibitem{ZJ12}
G.~Zhang and M.~R. James, ``Quantum feedback networks and control: a brief
  survey,'' {\em Chinese Science Bulletin}, vol.~57, no.~18, pp.~2200--2214,
  2012.

\bibitem{GZ15}
J.~E. Gough and G.~Zhang, ``On realization theory of quantum linear systems,''
  {\em Automatica}, vol.~59, pp.~139--151, 2015.

\bibitem{CKS17}
J.~Combes, J.~Kerckhoff, and M.~Sarovar, ``The {SLH} framework for modeling
  quantum input-output networks,'' {\em Advances in Physics: X}, vol.~2, no.~3,
  pp.~784--888, 2017.

\bibitem{ZLW+17}
J.~Zhang, Y.-x. Liu, R.-B. Wu, K.~Jacobs, and F.~Nori, ``Quantum feedback:
  theory, experiments, and applications,'' {\em Physics Reports}, vol.~679,
  pp.~1--60, 2017.

\bibitem{NY17}
H.~I. Nurdin and N.~Yamamoto, {\em Linear Dynamical Quantum Systems - Analysis,
  Synthesis, and Control}.
\newblock Springer-Verlag Berlin, 2017.

\bibitem{ZGPG18}
G.~Zhang, S.~Grivopoulos, I.~R. Petersen, and J.~E. Gough, ``The {K}alman
  decomposition for linear quantum systems,'' {\em IEEE Transactions on
  Automatic Control}, vol.~63, no.~2, pp.~331--346, 2018.

\bibitem{GC85}
C.~W. Gardiner and M.~J. Collett, ``Input and output in damped quantum systems:
  Quantum stochastic differential equations and the master equation,'' {\em
  Physical Review A}, vol.~31, no.~6, p.~3761, 1985.

\bibitem{BLP+90}
K.~Blow, R.~Loudon, S.~J. Phoenix, and T.~Shepherd, ``Continuum fields in
  quantum optics,'' {\em Physical Review A}, vol.~42, no.~7, p.~4102, 1990.

\bibitem{FKS10}
S.~Fan, S.~E. Kocabas, and J.~T. Shen, ``Input-output formalism for few-photon
  transport in one-dimensional nanophotonic waveguides coupled to a qubit,''
  {\em Physical Review A}, vol.~82, p.~063821, 2010.

\bibitem{FTR+18}
K.~A. Fischer, R.~Trivedi, V.~Ramasesh, I.~Siddiqi, and J.~Vu{\v{c}}kovi{\'c},
  ``Scattering into one-dimensional waveguides from a coherently-driven
  quantum-optical system,'' {\em Quantum}, vol.~2, p.~69, 2018.

\bibitem{TNP+11}
N.~Tezak, A.~Niederberger, D.~S. Pavlichin, G.~Sarma, and H.~Mabuchi,
  ``Specification of photonic circuits using quantum hardware description
  language,'' {\em Philosophical Transactions of the Royal Society A:
  Mathematical, Physical and Engineering Sciences}, vol.~370, no.~1979,
  pp.~5270--5290, 2012.

\bibitem{QPB+15}
Z.~Qin, A.~S. Prasad, T.~Brannan, A.~MacRae, A.~Lezama, and A.~Lvovsky,
  ``Complete temporal characterization of a single photon,'' {\em Light:
  Science \& Applications}, vol.~4, no.~6, pp.~e298--e298, 2015.

\bibitem{WM08}
D.~F. Walls and G.~J. Milburn, {\em Quantum Optics}.
\newblock Springer Science \& Business Media, 2007.

\bibitem{RL00}
R.~Loudon, {\em The Quantum Theory of Light}.
\newblock OUP Oxford, 2000.

\bibitem{YJ14}
N.~Yamamoto and M.~R. James, ``Zero-dynamics principle for perfect quantum
  memory in linear networks,'' {\em New Journal of Physics}, vol.~16, no.~7,
  p.~073032, 2014.

\bibitem{BR04}
H.-A. Bachor and T.~C. Ralph, {\em A Guide to Experiments in Quantum Optics}.
\newblock Wiley, 2004.

\bibitem{AL15}
A.~I. Lvovsky, ``Squeezed light,'' {\em Photonics: Scientific Foundations,
  Technology and Applications}, vol.~1, pp.~121--163, 2015.

\bibitem{GZ15b}
J.~E. Gough and G.~Zhang, ``Generating nonclassical quantum input field states
  with modulating filters,'' {\em EPJ Quantum Technology}, vol.~2, pp.~2--15,
  2015.

\bibitem{DZA16}
Z.~Dong, L.~Cui, G.~Zhang, and H.~Fu, ``Wigner spectrum and coherent feedback
  control of continuous-mode single-photon {F}ock states,'' {\em Journal of
  Physics A: Mathematical and Theoretical}, vol.~49, no.~43, p.~435301, 2016.

\bibitem{TG66}
U.~Titulaer and R.~Glauber, ``Density operators for coherent fields,'' {\em
  Physical Review}, vol.~145, no.~4, p.~1041, 1966.

\bibitem{RW20}
M.~G. Raymer and I.~A. Walmsley, ``Temporal modes in quantum optics: then and
  now,'' {\em Physica Scripta}, vol.~95, no.~6, p.~064002, 2020.

\bibitem{milburn08}
G.~J. Milburn, ``Coherent control of single photon states,'' {\em The European
  Physical Journal Special Topics}, vol.~159, pp.~113--117, Jun 2008.

\bibitem{Hassani13}
S.~Hassani, {\em Mathematical Physics: a Modern Introduction to its
  Foundations}.
\newblock Springer Science \& Business Media, 2013.

\bibitem{ZJ13}
G.~Zhang and M.~R. James, ``On the response of quantum linear systems to single
  photon input fields,'' {\em IEEE Transactions on Automatic Control}, vol.~58,
  no.~5, pp.~1221--1235, 2013.

\bibitem{GJN10a}
J.~E. Gough, M.~James, and H.~Nurdin, ``Squeezing components in linear quantum
  feedback networks,'' {\em Physical Review A}, vol.~81, no.~2, p.~023804,
  2010.

\bibitem{ZJ11}
G.~Zhang and M.~R. James, ``Direct and indirect couplings in coherent feedback
  control of linear quantum systems,'' {\em IEEE Trans. Automat. Contr.},
  vol.~56, pp.~1535--1550, 2011.

\bibitem{ZPL20}
G.~Zhang, I.~R. Petersen, and J.~Li, ``Structural characterization of linear
  quantum systems with application to back-action evading measurement,'' {\em
  IEEE Transactions on Automatic Control}, vol.~65, no.~7, pp.~3157--3163,
  2020.

\bibitem{JNP08}
M.~R. James, H.~I. Nurdin, and I.~R. Petersen, ``${H}^\infty$ control of linear
  quantum stochastic systems,'' {\em IEEE Transactions on Automatic Control},
  vol.~53, no.~8, pp.~1787--1803, 2008.

\bibitem{NJD09}
H.~I. Nurdin, M.~R. James, and A.~C. Doherty, ``Network synthesis of linear
  dynamical quantum stochastic systems,'' {\em SIAM Journal on Control and
  Optimization}, vol.~48, no.~4, pp.~2686--2718, 2009.

\bibitem{WRL61}
W.~R. Le~Page, {\em Complex Variables and the Laplace Transform for Engineers}.
\newblock Courier Corporation, 1980.

\bibitem{Z14}
G.~Zhang, ``Analysis of quantum linear systems{'} response to multi-photon
  states,'' {\em Automatica}, vol.~50, no.~2, pp.~442--451, 2014.

\bibitem{Z17}
G.~Zhang, ``Dynamical analysis of quantum linear systems driven by
  multi-channel multi-photon states,'' {\em Automatica}, vol.~83, pp.~186--198,
  2017.

\bibitem{GJN+12b}
J.~E. Gough, M.~R. James, and H.~I. Nurdin, ``Quantum filtering for systems
  driven by fields in single photon states and superposition of coherent states
  using non-markovian embeddings,'' {\em Quantum information processing},
  vol.~12, no.~3, pp.~1469--1499, 2013.

\bibitem{BC17}
B.~Q. Baragiola and J.~Combes, ``Quantum trajectories for propagating {F}ock
  states,'' {\em Physical Review A}, vol.~96, no.~2, p.~023819, 2017.

\bibitem{SKH13}
H.~Song, K.~B. Kuntz, and E.~H. Huntington, ``Limitations on the quantum
  non-{G}aussian characteristic of {S}chr{\"o}dinger kitten state generation,''
  {\em New Journal of Physics}, vol.~15, no.~2, p.~023042, 2013.

\bibitem{B80}
V.~Belavkin, ``Quantum filtering of markov signals with white quantum noise,''
  {\em Elektronika}, vol.~25, pp.~1445--1453, 1980.

\bibitem{belavkin1989nondemolition}
V.~P. Belavkin, ``Nondemolition measurements, nonlinear filtering and dynamic
  programming of quantum stochastic processes,'' in {\em Modeling and Control
  of Systems}, pp.~245--265, Springer, 1989.

\bibitem{PK98}
M.~B. Plenio and P.~L. Knight, ``The quantum-jump approach to dissipative
  dynamics in quantum optics,'' {\em Rev. Mod. Phys.}, vol.~70, pp.~101--144,
  Jan 1998.

\bibitem{vHSM05}
R.~van Handel, J.~Stockton, and H.~Mabuchi, ``Feedback control of quantum state
  reduction,'' {\em IEEE Transactions on Automatic Control}, vol.~50, no.~6,
  pp.~768--780, 2005.

\bibitem{bouten2007introduction}
L.~Bouten, R.~van Handel, and M.~R. James, ``An introduction to quantum
  filtering,'' {\em SIAM Journal on Control and Optimization}, vol.~46, no.~6,
  pp.~2199--2241, 2007.

\bibitem{barchielli2009quantum}
A.~Barchielli and M.~Gregoratti, {\em Quantum Trajectories and Measurements in
  Continuous Time: The Diffusive Case}, vol.~782.
\newblock Springer, 2009.

\bibitem{RR15b}
P.~Rouchon and J.~F. Ralph, ``Efficient quantum filtering for quantum feedback
  control,'' {\em Physical Review A}, vol.~91, no.~1, p.~012118, 2015.

\bibitem{DSC17}
A.~Dabrowska, G.~Sarbicki, and D.~Chru{\'s}ci{\'n}ski, ``Quantum trajectories
  for a system interacting with environment in a single-photon state: Counting
  and diffusive processes,'' {\em Physical Review A}, vol.~96, no.~5,
  p.~053819, 2017.

\bibitem{jug18}
J.~E. Gough, ``An introduction to quantum filtering,'' {\em arXiv preprint
  arXiv:1804.09086}, 2018.

\bibitem{GZP19}
Q.~Gao, G.~Zhang, and I.~R. Petersen, ``An exponential quantum projection
  filter for open quantum systems,'' {\em Automatica}, vol.~99, pp.~59--68,
  2019.

\bibitem{GZP20}
Q.~Gao, G.~Zhang, and I.~R. Petersen, ``An improved quantum projection
  filter,'' {\em Automatica}, vol.~112, p.~108716, 2020.

\bibitem{D20}
A.~M. Dabrowska, ``From a posteriori to a priori solutions for a two-level
  system interacting with a single-photon wavepacket,'' {\em JOSA B}, vol.~37,
  no.~4, pp.~1240--1248, 2020.

\bibitem{DZA18}
Z.~Dong, G.~Zhang, and N.~H. Amini, ``Single-photon quantum filtering with
  multiple measurements,'' {\em International Journal of Adaptive Control and
  Signal Processing}, vol.~32, no.~3, pp.~528--546, 2018.

\bibitem{serafini2004minimum}
A.~Serafini, S.~De~Siena, F.~Illuminati, and M.~G. Paris, ``Minimum decoherence
  cat-like states in {G}aussian noisy channels,'' {\em Journal of Optics B:
  Quantum and Semiclassical Optics}, vol.~6, no.~6, p.~S591, 2004.

\bibitem{BR08}
A.~M. Bra{\'n}czyk and T.~Ralph, ``Teleportation using squeezed single
  photons,'' {\em Physical Review A}, vol.~78, no.~5, p.~052304, 2008.

\bibitem{neergaard2013quantum}
J.~S. Neergaard-Nielsen, Y.~Eto, C.-W. Lee, H.~Jeong, and M.~Sasaki, ``Quantum
  tele-amplification with a continuous-variable superposition state,'' {\em
  Nature Photonics}, vol.~7, no.~6, pp.~439--443, 2013.

\bibitem{GEP+98}
K.~Gheri, K.~Ellinger, T.~Pellizzari, and P.~Zoller, ``Photon-wavepackets as
  flying quantum bits,'' {\em Fortschr. Phys.}, vol.~46, pp.~401--415, 1998.

\bibitem{FT20}
C.~Fabre and N.~Treps, ``Modes and states in quantum optics,'' {\em Reviews of
  Modern Physics}, vol.~92, no.~3, p.~035005, 2020.

\bibitem{ZWD+20}
H.-S. Zhong, H.~Wang, Y.-H. Deng, M.-C. Chen, L.-C. Peng, Y.-H. Luo, J.~Qin,
  D.~Wu, X.~Ding, Y.~Hu, {\em et~al.}, ``Quantum computational advantage using
  photons,'' {\em Science}, vol.~370, no.~6523, pp.~1460--1463, 2020.

\bibitem{ATF+21}
W.~Asavanant, K.~Takase, K.~Fukui, M.~Endo, J.-i. Yoshikawa, and A.~Furusawa,
  ``Wave-function engineering via conditional quantum teleportation with a
  non-gaussian entanglement resource,'' {\em Physical Review A}, vol.~103,
  no.~4, p.~043701, 2021.

\bibitem{RGM+03}
T.~C. Ralph, A.~Gilchrist, G.~J. Milburn, W.~J. Munro, and S.~Glancy, ``Quantum
  computation with optical coherent states,'' {\em Physical Review A}, vol.~68,
  no.~4, p.~042319, 2003.

\bibitem{OJT+07}
A.~Ourjoumtsev, H.~Jeong, R.~Tualle-Brouri, and P.~Grangier, ``Generation of
  optical ‘{S}chr{\"o}dinger cats’ from photon number states,'' {\em
  Nature}, vol.~448, no.~7155, pp.~784--786, 2007.

\bibitem{GGC+10}
T.~Gerrits, S.~Glancy, T.~S. Clement, B.~Calkins, A.~E. Lita, A.~J. Miller,
  A.~L. Migdall, S.~W. Nam, R.~P. Mirin, and E.~Knill, ``Generation of optical
  coherent-state superpositions by number-resolved photon subtraction from the
  squeezed vacuum,'' {\em Physical Review A}, vol.~82, no.~3, p.~031802, 2010.

\bibitem{HLR+15}
K.~Huang, H.~Le~Jeannic, J.~Ruaudel, V.~Verma, M.~Shaw, F.~Marsili, S.~Nam,
  E.~Wu, H.~Zeng, Y.-C. Jeong, {\em et~al.}, ``Optical synthesis of
  large-amplitude squeezed coherent-state superpositions with minimal
  resources,'' {\em Physical review letters}, vol.~115, no.~2, p.~023602, 2015.

\bibitem{EBK+15}
J.~Etesse, M.~Bouillard, B.~Kanseri, and R.~Tualle-Brouri, ``Experimental
  generation of squeezed cat states with an operation allowing iterative
  growth,'' {\em Physical review letters}, vol.~114, no.~19, p.~193602, 2015.

\bibitem{SUP+17}
D.~V. Sychev, A.~E. Ulanov, A.~A. Pushkina, M.~W. Richards, I.~A. Fedorov, and
  A.~I. Lvovsky, ``Enlargement of optical {S}chr{\"o}dinger's cat states,''
  {\em Nature Photonics}, vol.~11, no.~6, p.~379, 2017.

\bibitem{OJ18}
C.~Oh and H.~Jeong, ``Efficient amplification of superpositions of coherent
  states using input states with different parities,'' {\em JOSA B}, vol.~35,
  no.~11, pp.~2933--2939, 2018.

\bibitem{ENP19}
M.~Eaton, R.~Nehra, and O.~Pfister, ``Non-{G}aussian and
  {G}ottesman--{K}itaev--{P}reskill state preparation by photon catalysis,''
  {\em New Journal of Physics}, vol.~21, no.~11, p.~113034, 2019.

\bibitem{MPK+19}
E.~V. Mikheev, A.~S. Pugin, D.~A. Kuts, S.~A. Podoshvedov, and N.~B. An,
  ``Efficient production of large-size optical {S}chr{\"o}dinger cat states,''
  {\em Scientific Reports}, vol.~9, no.~1, pp.~1--15, 2019.

\bibitem{TYA+21}
K.~Takase, J.-i. Yoshikawa, W.~Asavanant, M.~Endo, and A.~Furusawa,
  ``Generation of optical {S}chr{\"o}dinger cat states by generalized photon
  subtraction,'' {\em Physical Review A}, vol.~103, no.~1, p.~013710, 2021.

\bibitem{DCZ20}
Z.~Dong, W.~Cui, and G.~Zhang, ``On the dynamics of a quantum coherent feedback
  network of cavity-mediated double quantum dot qubits,'' {\em
  arXiv:2004.03870}, 2020.

\bibitem{BB20}
T.~Breitenbach and A.~Borz{\`\i}, ``A sequential quadratic hamiltonian scheme
  for solving non-smooth quantum control problems with sparsity,'' {\em Journal
  of Computational and Applied Mathematics}, vol.~369, p.~112583, 2020.

\bibitem{CB16}
G.~Ciaramella and A.~Borzi, ``Quantum optimal control problems with a sparsity
  cost functional,'' {\em Numerical Functional Analysis and Optimization},
  vol.~37, no.~8, pp.~938--965, 2016.

\bibitem{BCS17}
A.~Borz{\`\i}, G.~Ciaramella, and M.~Sprengel, {\em Formulation and Numerical
  Solution of Quantum Control Problems}.
\newblock SIAM, 2017.

\end{thebibliography}

\end{document}